\documentclass[letterpaper,twocolumn,10pt]{article}
\usepackage{usenix-2020-09}

\usepackage{amsmath,amssymb,amsfonts}
\usepackage{amssymb}
\usepackage{amsthm}
\usepackage[noend]{algpseudocode}
\usepackage[ruled,vlined,linesnumbered]{algorithm2e}
\usepackage{booktabs}  
\usepackage{hyperref}
\usepackage{comment}
\usepackage{subcaption}
\usepackage{multirow}
\usepackage{multicol}
\usepackage{mathtools}
\usepackage{tikz}
\usepackage{cite}
\usepackage{graphicx}
\usepackage{textcomp}
\usepackage{xcolor}
\usepackage[inline]{enumitem}
\usepackage{pgf}
\usepackage{layouts}

\newtheorem{theorem}{Theorem}[section]
\newtheorem{definition}{Definition}
\newtheorem{lemma}[theorem]{Lemma}

\graphicspath{{images/}}

\begin{document}


\title{Isolated Scheduling for Distributed Training Tasks in GPU Clusters}


\author{Xinchi Han\textsuperscript{1}, Weihao Jiang\textsuperscript{1}, Peirui Cao\textsuperscript{1}, Qinwei Yang\textsuperscript{1},\\ Yunzhuo Liu\textsuperscript{1}, Shuyao Qi\textsuperscript{1}, Shengkai Lin\textsuperscript{1}, Shizhen Zhao\textsuperscript{1}\thanks{Corresponding author}\\
\emph{\textsuperscript{1}Shanghai Jiao Tong University}}


\maketitle

\begin{abstract}
Distributed machine learning (DML) technology makes it possible to train large neural networks in a reasonable amount of time. Meanwhile, as the computing power grows much faster than network capacity, network communication has gradually become the bottleneck of DML. Current multi-tenant GPU clusters face network contention caused by hash-collision problem which not only further increases the overhead of communication, but also creates unfairness and affects the user experience. In this paper, we firstly analyse how network contention affects the training time in a cluster with 32 NVIDIA V100 GPUs. Then we propose vClos to eliminate network contention by jointly optimizing network topology and communication pattern in distributed training. An OCS-vClos which introduces a layer of optical circuit switches (OCSs) in the leaf-spine network is also proposed to reduce potential network resource fragmentation caused by resource allocation strategy in vClos. Testbed experiments and real-trace-based large-scale simulations are conducted to demonstrate the superiority of vClos over existing network resource scheduling strategies.

\end{abstract}



\makeatletter

\section{Introduction}
\begin{figure}[htbp]
\centering
\input{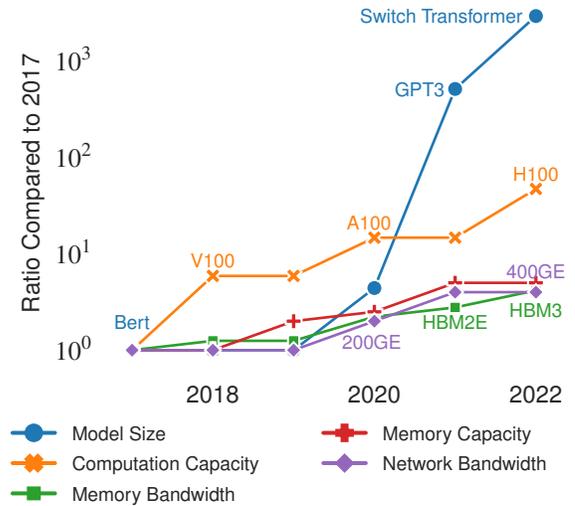}
\caption{The growth rate of network capacity is much lower than the growth rate of computing capacity and model size.}
\label{fig:NetworkGrowth}
\end{figure}

The field of machine learning has experienced exponential growth overtime, leading to the emergence of Distributed Machine Learning (DML). As a result, the size of individual machine learning models has skyrocketed from 1.5B in 2020 to over 1000B at present. However, the expansion of network bandwidth has not kept pace with this growth, as illustrated in Fig. \ref{fig:NetworkGrowth}. In the past five years, the scale of machine learning models has increased by 2941 times, while single-chip computing power has only improved by 47 times. Consequently, more and more tasks now require multi-server training, which demands interconnected communications. Unfortunately, single-port bandwidth has merely increased by a factor of four, exacerbating the network bottleneck already present in AI training. Open source datasets \cite{hu2021characterization, weng2022mlaas} reveal that an increasing number of DML tasks now employ hundreds or even thousands of GPUs. Therefore, the network has gradually become the bottleneck in large-scale DML systems, with networking alone accounting for up to 90\% of total training time~\cite{luo2020plink}.

Distributed training typically involves four types of parallelism: data parallelism, model parallelism, pipeline parallelism, and Mixture-of-Experts (MoE) parallelism. The training process usually involves two phases: forward propagation and backward propagation. To avoid the network from becoming a bottleneck in DML tasks, many training frameworks cover AllReduce communication overhead with computation overhead during backward propagation phases. However, the network still becomes the limiting factor for several reasons.
\begin{enumerate*}
\item DML tasks such as NLP and Recommendation Systems require AlltoAll communication traffic during training. Compared to AllReduce, 
cover AlltoAll communication overhead is more difficult, especially in large model training where AlltoAll communication overhead can account for $20\sim50\%$ of total training time.
\item  It is challenging to fully utilize network bandwidth in DML \cite{zhang2020network}, and merely increasing bandwidth strategies may not adequately address the network bottleneck. 
\item Modern data centers often use ECMP as the load balancing strategy, which faces hash collision problem, leading to bandwidth competition among different tasks in a multi-tenant cluster, resulting in higher communication overhead.
\end{enumerate*}

This paper focuses on investigating how network contention affects communication overhead and how to address the problem. An experimental testbed cluster with 32 NVIDIA V100 GPUs is set up to determine the order of severity of the hash collision problem and how it impacts communication overhead for a single DML task. Based on the findings, a \emph{vClos} scheme that can address the hash collision problem in a multi-tenant GPU cluster is proposed. \emph{vClos} aims to boost network resource utilization through eliminating network contention by jointly considering topology, routing, communication pattern, and GPU assignment.

The challenges of the contention-free task placement problem are two folds. There are mainly two sources of contention for DML tasks: intra-task contention and inter-task contention. \begin{enumerate*}
\item [\textbf{1. Intra-task contention:}] The communication stage of a DML task may contain multiple phases, each with a different communication pattern. A straightforward approach to eliminate contention is to design different contention-free forwarding rules for different phases. However, updating forwarding rules in modern switches/routers takes at least tens of microseconds, which is slow and error-prone.
\item [\textbf{2. Inter-task contention:}] When a network link is shared by multiple DML tasks, these tasks may contend for network bandwidth. One way to eliminate inter-task contention is to reserve a link-disjoint sub-topology for each DML task. However, this approach may incur additional resource fragmentation, i.e. even if there are enough GPU resources, a DML task still cannot be initiated due to insufficient network resources. Resource fragmentation can be detrimental to GPU resource utilization and increase the task waiting time.
\end{enumerate*}

Large model training typically involves multiple types of communication traffic, such as a typical large model involves data parallelism, MoE parallelism, model parallelism, and pipeline parallelism. Although previous work\cite{wang2022topoopt} has discussed partitioning mesh topology for tasks, AlltoAll communication in MoE parallelism performs poorly on mesh topology. DGX SuperPOD\cite{DGX2023} also recommended to use the Leaf-Spine architecture, therefore, in this article, we consider partitioning clos sub-topology for each task.

\emph{vClos} scheme we proposed addresses the challenges above for DML tasks in a Leaf-Spine network (i.e. a folded Clos). Given a single DML task, \emph{vClos} is a smaller Leaf-Spine sub-topology and adopts a \emph{Source Routing} strategy which can be proved to be contention-free for any \emph{Leaf-wise Permutation Traffic Pattern} defined in Section \ref{leaf-wise}. By assigning GPUs properly, most communication patterns can be shaped to meet the pattern.

In a multi-tenant data center network, \emph{vClos} can adapt an optional layer of Optical Circuit Switches (OCS) to reduce resource fragmentation while guaranteeing network isolation among different DML jobs. The power consumption of OCSes are negligible. When a DML task cannot be scheduled normally due to insufficient network resources, \emph{OCS-vClos} can reconfigure the network topology by controlling the OCS layer to make the task deployable.

The key contributions are summarized below:
\begin{enumerate*}
\item Detailed analysis of how network contention affects DML training time cost through testbed experiments on a cluster with 32 V100 GPUs;
\item Proves network contention-free communication is feasible for most DML jobs that follow a \emph{Leaf-wise Permutation Traffic Pattern};
\item Proposed a feasible \emph{vClos} scheme that allocates an exclusive  sub-topology for each job using the link reservation policy;
\item In \emph{OCS-vClos} an additional OCS layer is introduced to reduce resource fragmentation caused by resource scheduling strategy in \emph{vClos}.
\end{enumerate*} The feasibility of the scheme is verified through a cluster containing 32 v100 GPUs, and its performance in large-scale networks is demonstrated through simulation.

\section{Background}
\subsection{With the development of large models, networks have gradually become potential bottlenecks}
With the rise of DML, computing demand of large Al models doubles every 3.4 months since 2012~\cite{web:demand_doble}. The BERT~\cite{devlin2018bert} model contains 300 million parameters, the GPT-2~\cite{budzianowski2019hello} model contains 1.5B parameters, and GPT-3~\cite{floridi2020gpt} model even consists of 175B parameters. According to NVIDIA's estimation, if GPT-3 is to be trained, even if the video/host memory can be installed, using 8 V100 GPUs (configured with a DGX-1 server) will take 36 years \cite{web:full_use_GPU}. 

In recent years, GPU computing power has also been continuously increasing, the computing power of a single GPU has increased from 21.2 TFLOPS on P100 to 1000 TFLOPS on H100. However, the growing speed of AI-model computing demands outpaces Moore’s Law, which brings challenges to the development of AI technology. The growth rate of the model is greater than the growth rate of GPU computing power, so the number of GPUs required is increasing, which boost the market demand for large-scale GPU clusters. 

However, the growth rate of network bandwidth is relatively slow compared with GPU computing power, and is far slower than the computing power demand of the model, so the bottleneck effect of the network is becoming increasingly apparent. The comparison of the growth rate between the network and other major cluster components in Fig. \ref{fig:NetworkGrowth} further illustrates this trend.

In order to meet the needs of training large-scale DNN models, distributed training comes to birth. Recent years, researchers have developed several useful frameworks for distributed training, including MXNet~\cite{chen2015mxnet}, TensorFlow~\cite{abadi2016tensorflow}, PyTorch~\cite{paszke2019pytorch}, Horovod~\cite{sergeev2018horovod}, MindSpore~\cite{tong2021study}, etc. Distributed training typically consists of two phases: computation and communication. In the computation phase, each worker calculates the gradients of the model parameters in \emph{forward and backward propagation} procedure~\cite{lecun1988theoretical} using \emph{stochastic gradient descent} algorithm; in the communication phase, all the workers synchronize all the gradients and then update the model parameters used in the next iteration. Distributed training is highly expensive and time-consuming. \cite{rashidi2022themis} shows that for tasks with high communication to computation ratio $\alpha$, network bandwidth is the primary factor that determines communication latency. In some extreme cases, 90\% of the training time can be spent on networking communication in distributed training~\cite{luo2020plink}. With the popularity of large model training \cite{vaswani2017attention, rashidi2022themis}, the need for high communication efficiency is increasing. 

Previous work\cite{zhang2020network} has mentioned that the network may be a bottleneck in DML due to the low utilization of the network caused by network protocol overhead, but there may be other factors as well. Different from \cite{zhang2020network}, our testbed experiment is based on a 100Gbps RoCE instead of TCP, and with GDR enabling to maximize communication efficiency. In this case, hash-collision problem can result in a noticeable communication overhead. In a multi-tenant GPU cluster, network contention can cause two additional problems. First, the longer communication time of a single task will increase the waiting time of subsequent tasks, especially when the cluster load rate is high. Second, users may purchase services with different quality at the same price due to the network environments, which will affect the user experience.


\section{Understand network contention for DML tasks}
\subsection{The severity of hash-collision problem in the network}
\label{hash-collision-chosen}
\begin{figure}[htbp]
\centering
\input{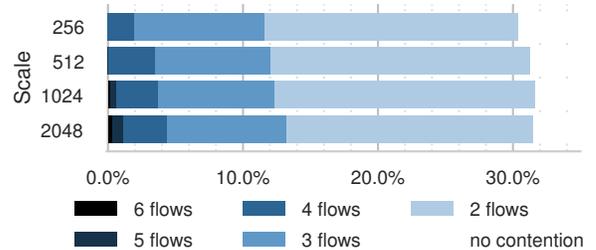}
\caption{The proportion of flow contention under different scales (total is 100\%)}\label{fig:ecmp_contention}
\end{figure}

\begin{figure}[htbp]
\centering
\def\svgwidth{0.5\linewidth}
\input{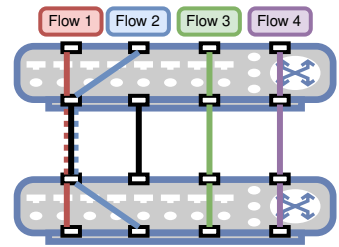}
\caption{An example of hash collision.}\label{fig:hash_collison}
\end{figure}

\begin{figure}[!htbp]
\centering
    \begin{subfigure}[b]{\linewidth}
         \centering
        \includegraphics[width=\linewidth]{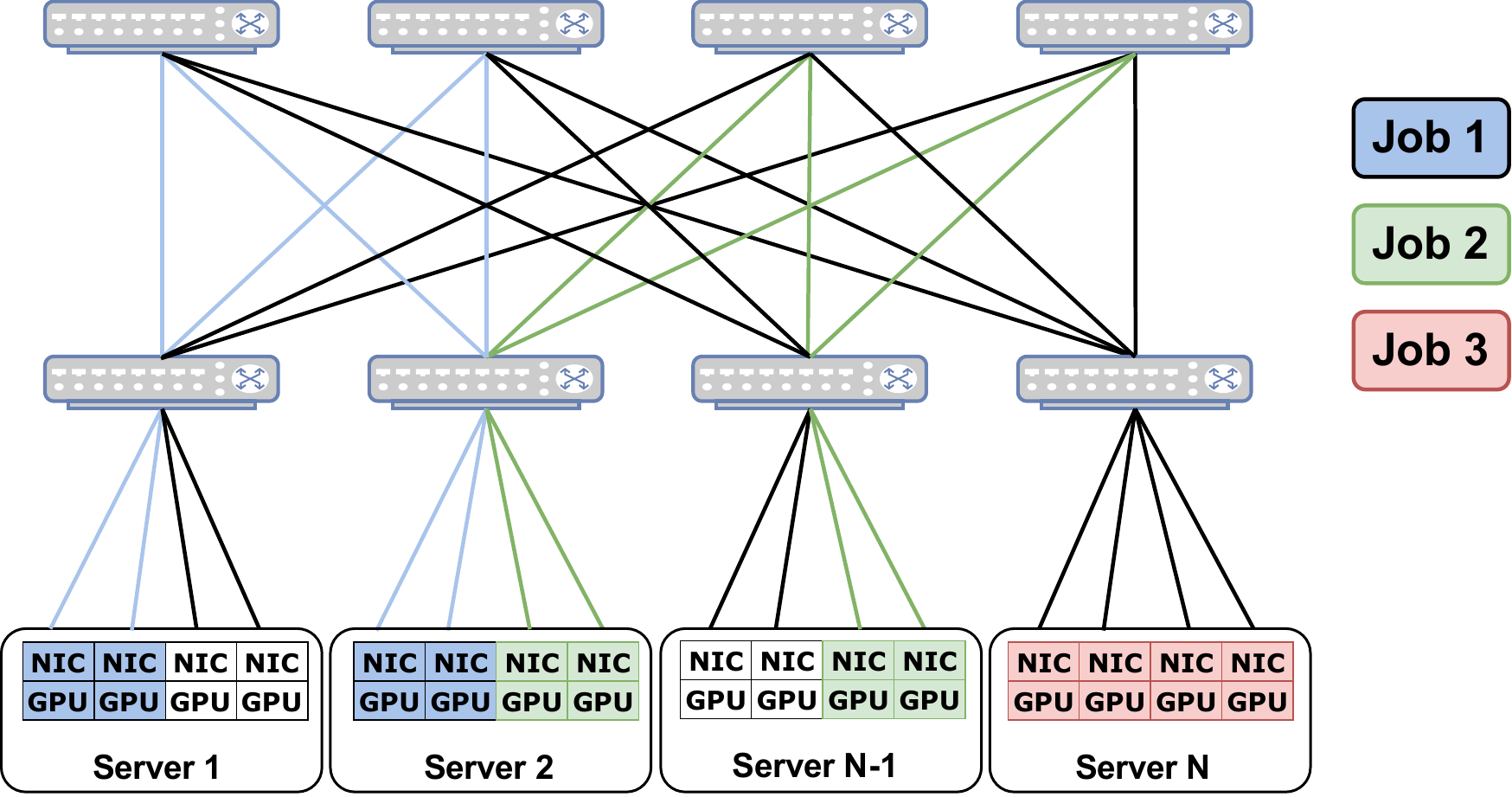}
         \caption{Example of virtual Clos in vClos}\label{fig:vClosExample}
     \end{subfigure}
     
    \begin{subfigure}[b]{\linewidth}
         \centering
         \includegraphics[width=\linewidth]{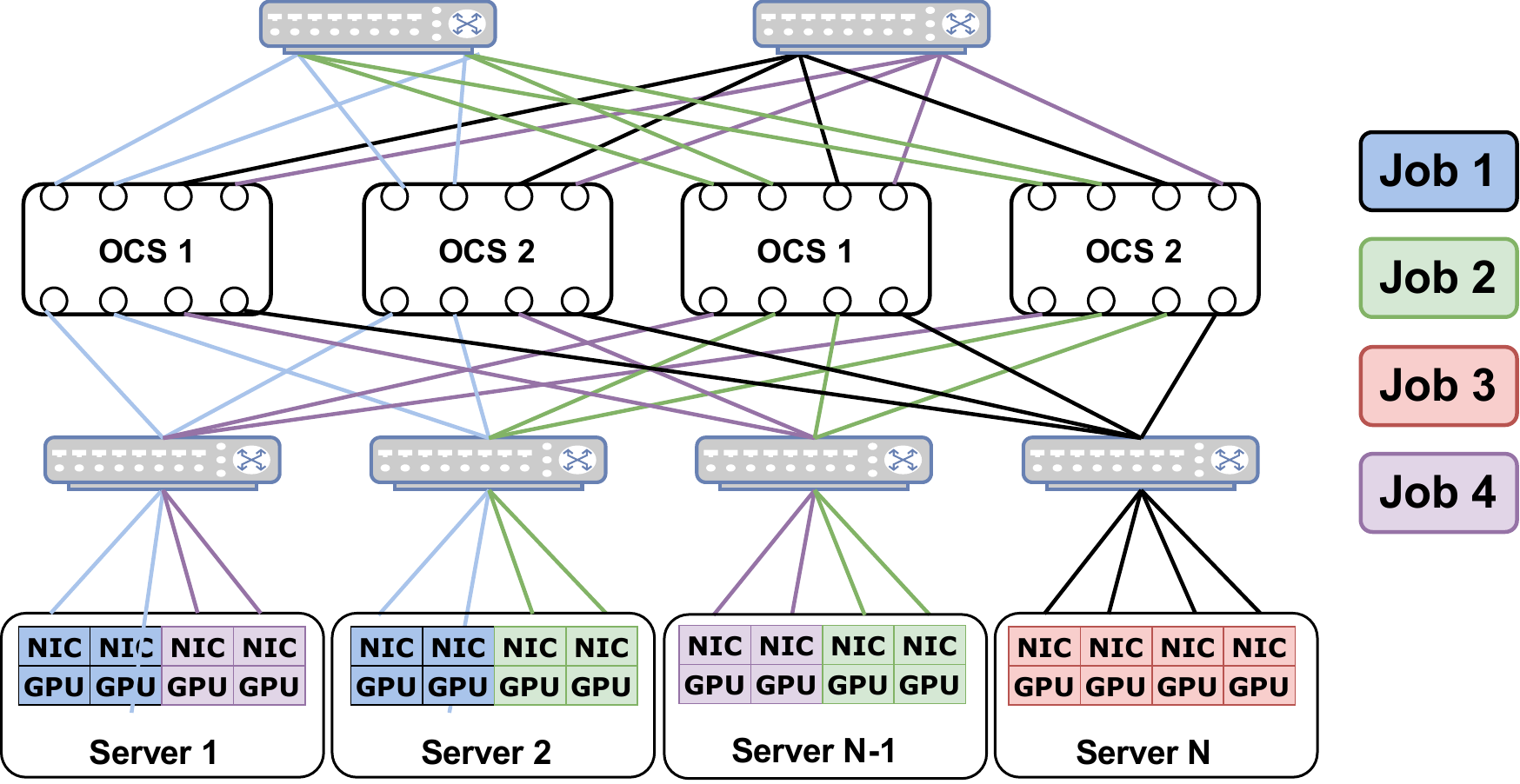}
         \caption{Example of virtual Clos in OCS-vClos}\label{fig:OCS-vClosExample}
     \end{subfigure}
\caption{vClos/OCS-vClos physical architecture.}
\end{figure}

Hash-collision is a common problem in modern data centers. Before measuring the impact of network collisions, we first need to know the probability of hash-collision occurring. Selecting the best combination from many hash factor and hash mode combinations is a common means of mitigating network contention. We conduct a testbed experiment based on two servers and two Huawei CE8850 switches as shown in Fig. \ref{fig:hash_collison}. We randomly select 4 GPUs in each server to communicate with each other to generate 20 Ring-AllReduce communication tasks, and select 15 hash-mode--hash-factor combinations to calculate the probability of hash-collision. However, no matter what hash mode-hash factor combination is chosen, hash-collision cannot be completely avoided. An example of hash collision is shown in Fig. \ref{fig:hash_collison}. Under the optimal combination of hash factors, there is still an average 31.5\% probability of network contention occurring, and even a 3\% probability of all flows routing to the same port. Simulation tests on a large-scale cluster is also conducted based on the Leaf-Spine architecture in Fig. \ref{fig:vClosExample} to demonstrate the severity of hash-collision at different scales. Mmh3 is chosen as the hash function and the 5-tuple \textsf{(src-ip,dst-ip,src-port,dst-port,port number)} is chosen as the hash factor. Fig. \ref{fig:ecmp_contention} shows that the larger the cluster size, the greater the likelihood of network contention. In extreme cases, there is a 3.3\% probability of 6 flows competing for bandwidth. So it is necessary to study the impact of hash-collision on large-scale GPU clusters.

\subsection{Scaling Factor of DML Tasks}
Scaling factor is often used to measure the scale-out ability, which is defined as Equation \ref{eq:scaling_factor}, where $T$ is the throughput using single device and $T_n$ is the throughput of a system with $n$ devices. The ideal scaling factor would be 1, and communication bottlenecks contribute most decrement in actual models and environment.

\begin{equation}
\label{eq:scaling_factor}
  ScalingFactor = \frac{T_n}{nT} 
\end{equation}

Previous work \cite{zhang2020network} has shown the scaling factor in a TCP environment. Compared with TCP, RoCE protocol result in lower communication overhead, and with GDR enabling can further improve communication efficiency, so a testbed experiment with 32 V100 GPU interconnected by a 100Gbps RoCE network is conducted, the network topology is the standard Leaf-Spine architecture in Fig. \ref{fig:vClosExample}.

We test two case: contention-free network and ECMP-based network. In the first case, we plan the route in advance by using ACL to ensure that there is no contention. In the second case, we select the 5-tuple \textsf{(src-ip,dst-ip,src-port,dst-port,port number)} as the hash factor and set the hash mode to 5 for testing. Fig. \ref{fig:scaling_factor} shows the changes of scaling factor for different models, where the bar chart shows the SF values when using ECMP, the line chart shows the SF values under different bandwidths without contention. 

When using ECMP, the scaling factor may decrease as the number of nodes increases, this may because the more nodes there are, the more severe the contention becomes. Fig. \ref{fig:scaling_factor} indicates for some tasks such as VGG16 or BERT, network contention can significantly affect the scaling factor. The noticeable impact of hash-collision on the scaling factor can be seen in 100GE networks which suggests that hash-collision in large-scale clusters may have a visible impact on training performance. It is worth noting that these results are tested on V100 GPUs, and if the cluster adopts stronger GPUs like H100, the impact of hash-collision on performance could be worse.

\begin{figure*}[!htb]
\centering
\input{images/scaling-factor.pgf}

{\footnotesize\textsf{Batch sizes: Bert = 4, VGG16, ResNet50, ResNet101 = 32}}
\caption{Scaling Factor may Decrement Due to Hash-collision.}
\label{fig:scaling_factor}
\end{figure*}

\subsection{How does network contention affect communication overhead for single DML task?}
In Fig.~\ref{fig:scaling_factor}, we note that different models have different sensitivities to network contention, a simple testbed experiment is conducted to illustrate this phenomenon, with a cluster consists of two machines each contains 4 V100, and bandwidth competition between the two flows is deliberately introduced through route planning.

Fig.~\ref{fig:contention_impact} shows the impact of two flows contention on training time. We observed that: \begin{enumerate*}
    \item For tasks as VGG16 and BERT, network contention may result in a obvious throughput decrement which may due to their relatively larger parameter number;
    \item For all data parallelism models, the larger the batch size, the smaller the impact of contention, which may because for data parallelism DML with larger batch sizes, computational overhead is more easier to be covered;
    \item Models such as DLRM and MoE are relatively sensitive to contention, which may because the AlltoAll communication is difficult to be covered;
    \item DML tasks running in smaller bandwidths are more sensitive to network contention, which may be due to two reasons: firstly, the non-coverable communication overhead increases in smaller bandwidths, and secondly, the synchronization overhead of communication algorithms prevents the network bandwidth from being fully utilized, which implies that the severity of network contention has a non-linear impact on the performance of DML training, so avoiding hash-collision to reduce the number of flow contention is meaningful.
\end{enumerate*}

\begin{figure}[ht]
\centering
\input{images/new_noncov.pgf}
\caption{Expert parallel and some recommendation models such as DLRM are more sensitive to network contention. In extreme cases, two flow contention can result in a throughput decrease of about 60\%}
\label{fig:contention_impact}
\end{figure}


\subsection{How does network contention affect multi-tenant Cluster}
The case for multi-tenant clusters is more complex than the single-task scenario. Naturally, contention-affected tasks will increase the average DML job running time (JRT). However, the impact of network contention on the average completion time (JCT) of tasks may be more pronounced. The reason is that when a task runs longer, the wait time of all subsequent waiting tasks also becomes longer, so the average job waiting time (JWT) will also increase. In addition, the JCT of tasks with the same parameters may fluctuate due to the network environment, as a result, the service quietly with the same price are different, which affects the user experience.

\section{Overview}
\subsection{Network Architecture}
\emph{vClos} is designed in a Leaf-Spine network architecture (Fig. \ref{fig:vClosExample}). The Leaf-Spine architecture has been widely adopted to build high-speed compute networks for GPU clusters \cite{DGX2023}. 

In the Leaf-Spine architecture, each server contains a fixed number of GPUs (e.g. 8) and these GPUs are internally connected via Nvidia's NVLinks and NVSwitches, which could offer a Tbps-level bandwidth for intra-server communication between each GPU pair. GPUs from different servers communicate through the network via Network Interface Cards (NIC). A modern NIC, e.g. Mellanox CX5 \cite{2021Breakfast}, could offer a communication bandwidth of 100/200Gbps. Since a server contains multiple GPUs, to avoid PCIE contention, we bound each GPU to a single NIC. This design was also adopted in Alibaba's GPU cluster EFLOPS~\cite{dong2020eflops}. To guarantee full bisection bandwidth, each Leaf switch has half of its ports connected to the server NICs and half of its ports connected to the Spine switches. The topology between all the Leaf switches and all the Spine switches is a uniform bipartite graph, with an equal number of links between every pair of Leaf-Spine switches. Such a Leaf-Spine architecture could support a small to medium scale GPU cluster. For example, if each Leaf or Spine switch has 64 ports, then the maximum number of GPUs is $32\times 64=2048$.

\subsection{Understanding the Communication Patterns of DML Tasks}
Common DML tasks involves one or more of the following parallel parallelism: data parallelism, model parallelism, pipeline parallelism, and expert parallelism. Different parallel paradigms have different communication patterns. 

\begin{figure*}[!htb]
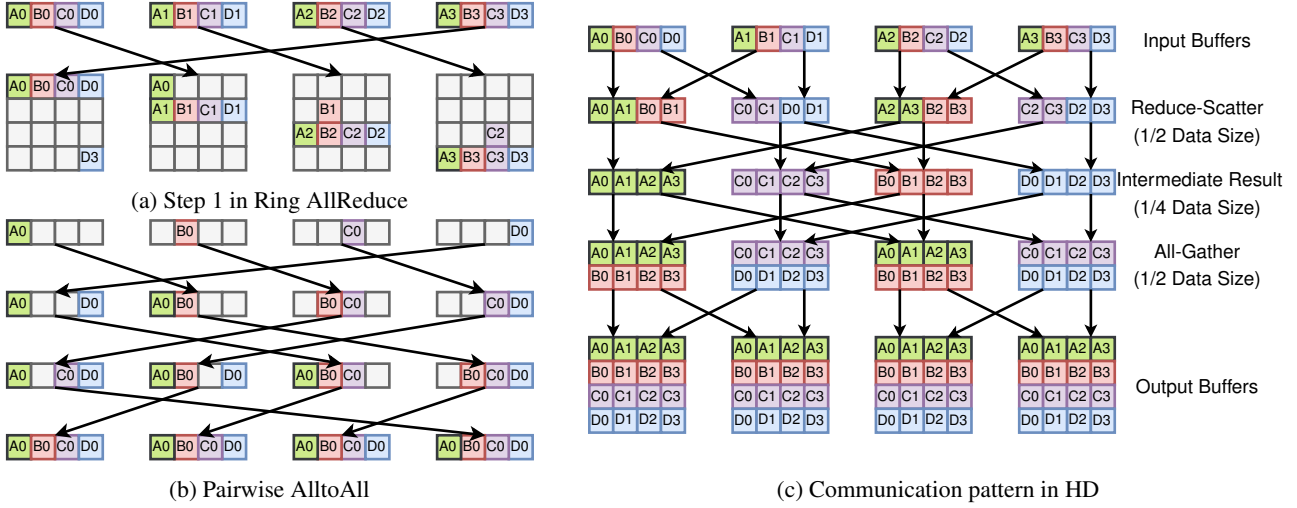

\centering
\hfill
\begin{subfigure}[b]{0.40\textwidth}
\centering
\input{images/ring-first-step.tikz}
\caption{Step 1 in Ring AllReduce}\label{fig:ring_pattern}
\vfill

\input{images/pairwise-all-to-all.tikz}
\caption{Pairwise AlltoAll}\label{fig:alltoall}
\end{subfigure}
\begin{subfigure}[b]{0.59\textwidth}%
\centering
\vfill
\input{images/hd.tikz}
\vspace{10px}
\caption{Communication pattern in HD}\label{fig:hd}
\end{subfigure}
\hfill

\caption{Communication Patterns}\label{fig:comm_pattern}
\end{figure*}

In data parallelism each GPU stores the same model and processes different part of data set, and GPUs need to synchronize parameters which is usually realized through AllReduce. Typical data parallelism DML tasks contains forward computation, backward computation and AllReduce communication. Common AllReduce communication algorithms include Ring-AllReduce (Ring) in Fig. \ref{fig:ring_pattern}, recursive half-doubling(HD) in Fig. \ref{fig:hd}, hierarchical ring-AllReduce and other algorithms. The data size of AllReduce flow is generally around a few megabytes or tens of megabytes, and most of the communication overhead can be covered by backward computation. However, communication overhead may not be well covered up when multiple flows competing for bandwidth.

In model parallelism each GPU store different parts of the same model. Different model segmentation methods may result in different communication characteristics. Communication in model parallelism cannot be covered by computation, and in order to efficiently utilize the higher bandwidth within the machine, model parallelism is usually executed within the machine\cite{fei2021efficient}. 

Pipeline parallelism\cite{fei2021efficient} improves parallelism by dividing a large batch into multiple small batches and synchronizing parameters after all small batches have been calculated. Only send/receive communication between neighbour GPUs exists in pipeline parallelism. Compared to data parallelism, pipeline parallelism has a smaller amount of communication size, but the communication overhead cannot be covered up by computation overhead.

Expert parallelism is widely used in large model training. A MoE layer exists in expert parallelism which consists of multiply experts placed on different workers. MoE layer requires All2All communication between different experts. Compared with data parallelism, the communication overhead in All2All can not be covered. The data size of All2All flows is generally around tens of megabytes, so expert parallelism is sensitive to network contention. 

A typical large model training process may involve multiple parallel methods mentioned above. We analyze the training overhead of a large model with 600 billion parameters without preventing network contention, in which All2All communication overhead accounting for 25.8\% and allreduce communication accounting for 4.2\%.


\subsection{Objective}
Since Leaf-Spine architecture is rearrangeable-nonblocking, it seems possible to plan routes based on different communication patterns to avoid contention, but there are two reasons that make this difficult to achieve:\begin{enumerate*}
    \item If modifying routing table affects the existing tasks, packet loss may occur and may result in PFC storms in large-scale networks;
    \item  It is difficult to directly perceive user traffic characteristics in a public cloud environment due to privacy considerations.
\end{enumerate*} Therefore, we hope to eliminate network contention without perceiving communication patterns. The objective of \emph{vClos} is to eliminate link contention for DML tasks in a multi-tenant GPU clusters. We focus on two scenarios in this paper.

\paragraph{Single Task Case} The common communication patterns of DML tasks is analyzed and network contention can be eliminated by designing a content free routing in Leaf-Spine networks. Specifically, in Leaf switch, a scheduler plan the route when some tasks arrives by issuing ACL flow table, and in Spine switch, the route table is directly generated through BGP or static routing. In this way, traffic contention within the task can be eliminated.

\paragraph{Multi-Tasks Case} Compared with single task case, multi-tasks scenario is more complex. Therefore, the scheduler attempt to partition  sub-topologys \emph{vClos} for each task in a Leaf Spine network to isolate traffic between tasks. In order to reduce potential network resource fragmentation, which means even if the cluster has sufficient GPU resources, tasks still need to wait due to the inability to find suitable network resources to partition \emph{vClos}, we also consider introducing a layer of OCS between Leaf switches and Spine switches to reduce fragmentation of network resources through OCS reconfiguration.

\begin{figure}[htbp]
\centering
\includegraphics[width=\linewidth]{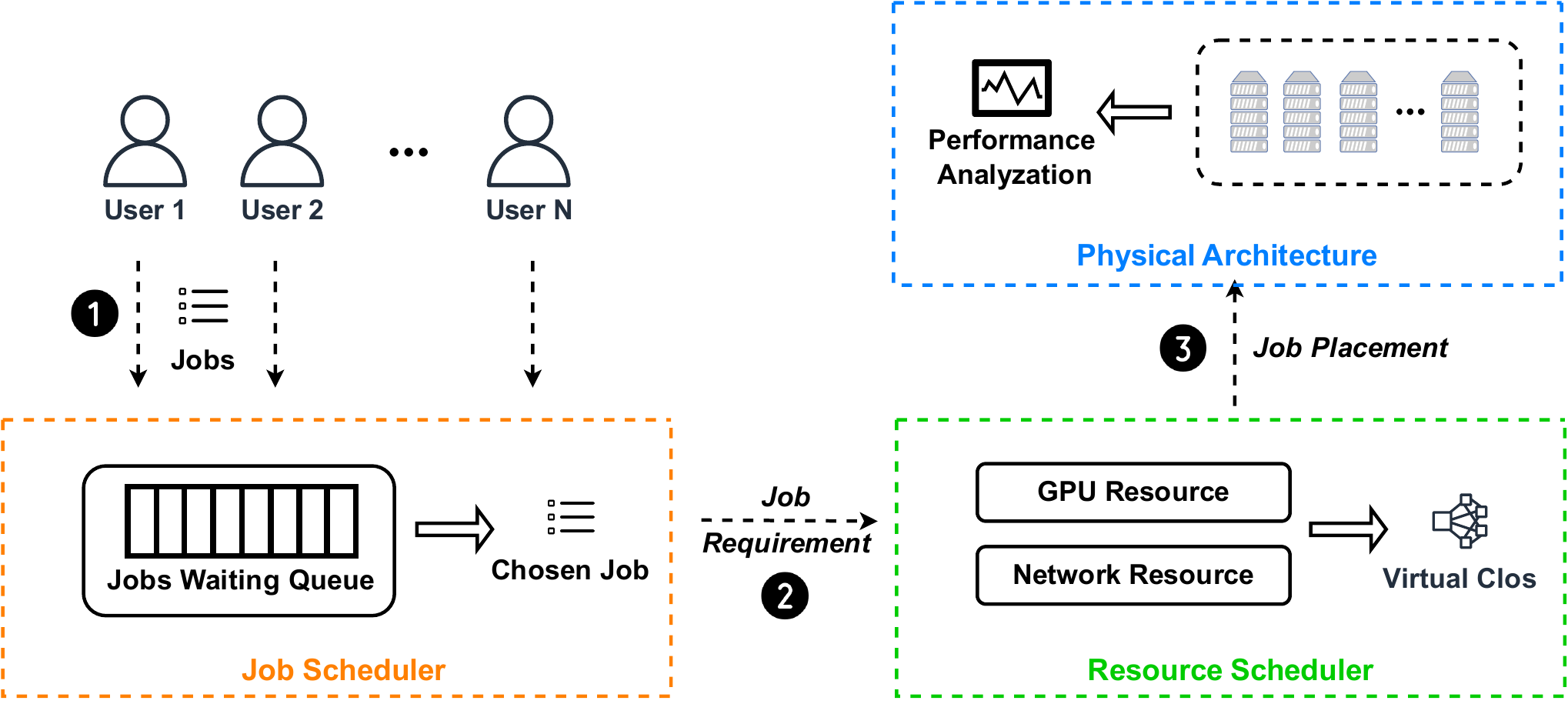}
\caption{Overview of Isolated Scheduling}
\label{workload_flow}
\end{figure}

\paragraph{Overview} Fig. \ref{workload_flow} shows an overview of proposed Isolated scheduling. The DML task first enter a waiting queue after it is submitted, and the task scheduler will select a task from the waiting queue and attempt to allocate resources. The most commonly used task scheduling algorithm in the industry is First-in-First-out (FIFO), and this paper mainly considers the case of FIFO. We also conducted experiments on the cases of Smallest-First(SF) and Early-Deadline-First(EDF) in Section \ref{job_scheduler} to analyse the sensitivity to task scheduling. Resource Scheduler is the key of this paper, which assign GPU resource, issue flow tables to the switch to plan the route, and configure the VRF command to generate virtual Leaf/virtual Spine to generate \emph{vClos} for each chosen job. Performance indicators such as task running time and waiting time are also monitored to evaluate the performance of different solutions.

\section{Preliminaries to Eliminate Network Contention}
\subsection{Difficulty of Eliminating Contention}
We aim to design a contention-free resource scheduler. Since the Leaf-Spine architecture is rearrangeably nonblocking, any permutation traffic pattern can be routed in the Leaf-Spine architecture without link contention. As we will see shortly, the communication pattern of the collective-communication based distributed training tasks belongs to the category of permutation patterns. Hence, if we were able to update the switch forwarding table as soon as the communication pattern changes, then we can completely eliminate network link contention. However, this approach incurs significant control overhead. First, computing a feasible set of contention-free forwarding rules for a given permutation pattern in a rearrangeably-nonblocking Leaf-Spine architecture takes time. Second, updating switch forwarding tables can be error-prone and may interrupt the on-going flows. Third, the communication patterns of some distributed training tasks may change on the order of tens of microseconds. Even though the detailed communication patterns of distributed training tasks are predictable, switch forwarding rule update cannot be accomplished in such short amount of time.

Another strategy to achieve contention-free is to build a strictly-nonblocking Leaf-Spine architecture. This approach stills creates a uniform bipartite graph between all the Leaf switches and all the Spine switches, but the number of uplinks $p_u$ of each Leaf switch and the number of downlinks $p_d$ of each Leaf switch must satisfy $p_u\geq 2*p_d - 1$. Then, when a new flow arrives, there must exist a Spine switch, such that both the link from the source Leaf to this Spine switch and the link from this Spine switch to the destination Leaf are not occupied. Using this strategy, it is easy to compute forwarding rules for new flows, and programming these new forwarding rules to switches does not affect the on-going flows. However, this architecture still faces the following problems. First, adding new forwarding rules takes non-negligible amount of time. Second, the strictly-nonblocking Leaf-Spine architecture is less efficient than the rearrangeably-nonblocking Leaf-Spine architecture. For example, using 64-port switches, we can build a rearrangeably-nonblocking Leaf-Spine network with $32\times 64=2048$ GPUs using $32+64=96$ switches. In contrast, using 64-port switches, we can only build a strictly-nonblocking Leaf-Spine network with $21\times 64=1344$ GPUs using $(21\times 2 - 1)+64=105$ switches.

\subsection{Contention-Free Routing in Leaf-Spine Networks}
\label{leaf-wise}
In this section, a sufficient condition that guarantees contention free in Leaf-Spine networks is provided. The results in this section will motivate the design of \emph{vClos}.

Consider the following routing strategy. Note that each Leaf switch has an equal number of server-facing ports and spine-facing ports. In the $m$-th Leaf switch $\mathcal{L}_m$, we create an arbitrary one-to-one mapping $f_m(\cdot)$ from the server-facing ports to the spine-facing ports, such that no two server-facing ports map to the same spine-facing port. Then, we can route a packet from a source GPU $\mathcal{G}_s$ to a destination GPU $\mathcal{G}_d$ based on the following strategy:
\begin{description}
\item[Case 1:] $s$ and $d$ connect to the same Leaf switch. We can directly forward the packet from $s$ to $d$.
\item[Case 2:] $s$ and $d$ connect to different Leaf switches. Assume that $s$ connects to the $i$-th server-facing port of the $m$-th Leaf switch. We will first forward the packet to the $f_m(i)$-th spine-facing port of the $m$-th Leaf switch, and then from that Spine switch,  we forward the packet to the destination. (Note that there is an unique path from any Spine switch to any GPU.)
\end{description}

In a Leaf-Spine network, any Spine switch can be used to forward a packet from its source to its destination. Note that in the above routing strategy, it is the source that determines the route a packet takes. Hence, in the rest of this paper, we call the above routing strategy as \emph{Source Routing}.

\begin{definition}[Leaf-wise Permutation Traffic Pattern]\label{def:perm_traffic_pattern}
Assume that there are $N$ servers and $L$ Leafs in the network. Any permutation $(G_0,G_1,\dots,G_{N-1})$ of $(0,1,\dots,N-1)$ defines a permutation pattern: the $s$-th GPU $G_i$ only sends traffic to the $d$-th GPU $G_{d}$. The above permutation pattern is called a Leaf-wise permutation pattern, if there exist another permutation $(l_0,l_1,\dots,l_{L-1})$ of $(0,1,\dots,L-1)$, such that any flow from $G_i$ to $G_{d}$ satisfies one of the following two conditions: 1) $G_i$ and $G_{d}$ belong to the same Leaf switch; 2) there exists $j$, such that $G_i$ directly connects to the $j$-th Leaf $l_j$ and $G_{d}$ directly connects to the ${j^*}$-th Leaf $l_{j^*}$ where $j\neq {j^*}$. For any $G_i$ and $G_{s^*}$, if $G_i$ and $G_{s^*}$ directly connects to different Leaf switches, the $G_{d}$ and $G_{d^*}$ directly connects to different Leaf switches. 
\end{definition}

A source routing strategy can be completely determined by $L$ mapping functions $(f_0(\cdot), f_1(\cdot), \dots, f_{L-1}(\cdot))$. For any source routing strategy, we have the following lemma.

\begin{lemma}\label{lem:network_contention}
A source routing strategy can guarantee contention-free for any Leaf-wise permutation traffic pattern.
\end{lemma}

\begin{proof}
If $G_i$ and $G_{d}$ directly connect to the same Leaf switch, the proposed source routing strategy  can naturally  guarantee contention-free, we prove another case by contradiction, where two flows cross Leafs competes bandwidth. Suppose there exists permutation $(s_0,s_1,\dots,s_{S-1})$ of $(0,1,\dots,S-1)$, such that $G_i$ directly connects to $j$-th Leaf $l_j$ and $G_{d}$ directly connects to the ${k}$-th Leaf $l_{k}$ where $j\neq k$. $l_j$ connect to the $j$-th port of $m$-th Spine switch $s_m$ and $l_{k}$ connect to the ${k}$-th port of $s_m$.

Suppose there exists another pair of GPUs $G_{i^*}$ and $G_{d^*}$ competes bandwidth with $G_i$ and $G_{d}$. $G_{i^*}$ directly connects to $j^*$-th Leaf $l_{j^*}$ and $G_{d^*}$ directly connects to the ${k^*}$-th Leaf $l_{k^*}$ where ${j^*}\neq{k^*}$. $l_{j^*}$ connect to the ${j^*}$-th port of $m^*$-th Spine switch $s_m^*$ and $l_{k^*}$ connect to the ${k^*}$-th port of $s_m^*$.

There are two cases when network contention happens: 1) $j = {j^*}$, where $G_i$ and $G_{i^*}$ directly connects to the same Leaf but send data to same Spine switch, which is contrary to source routing. 2) ${k} = {k^*}$ and ${t} = {t^*}$, which means two GPUs aim at the same Leaf are routed through the same Spine, which is contrary to Definition \ref{def:perm_traffic_pattern}. 


\end{proof}

\subsection{Eliminate Network Contention for a Single Distributed Training Task in a Leaf-Spine Network}
We analyzed the communication algorithms often involved in GPU cluster in recent years\cite{chu2020nv,sergeev2018horovod,wu2023sip,paszke2019pytorch,wang2022topoopt,sanghoon2023logical,awan2019scalable,dryden2018aluminum,ueno2019exhaustive,xu2022ring,jia2018highly,jiang2020unified,jeaugey2019massively,thangakrishnan2020herring}, and found that most of the communication algorithms follow the traffic pattern in Definition \ref{def:perm_traffic_pattern}. We consider a case where GPU $G_i$ sends send flow to $G_d$, the communication path can be represented as $G_i \rightarrow l_j \rightarrow s_m \rightarrow l_k \rightarrow G_d$. We will show that for these communication algorithms for any other communication path  $G_{i^*} \rightarrow l_{j^*} \rightarrow s_{m^*} \rightarrow l_{k^*} \rightarrow G_{d^*}$, if $j!=j^*$ and $m=m^*$, then $k!=K^*$.

\noindent\textbf{Data parallelism}: With data parallelism, each GPU stores a copy of the whole model and handle different part of the dataset. In each iteration, GPUs aggregate their gradients locally and synchronize weights through communication algorithms such as ring allreduce. There are in general two paradigms for data parallelism: 1) parameter server~\cite{thangakrishnan2020herring} and 2) collective communication~\cite{wu2023sip,rashidi2022themis,romero2022accelerating,wang2020blink}. In the parameter server mode, many workers may send gradients to the same parameter server at the same time, thus creating a communication bottleneck. Collective communication could evenly distribute the communication workload among all the workers, and thus can be more efficient for distributed training.  Using Lemma \ref{lem:network_contention}, we show that many collective-communication based distributed training tasks can be contention-free if properly deployed. 
AllReduce is one of the most frequently used collective communication procedures in distributed training. AllReduce performs a global reduction operation (e.g., SUM, AVERAGE, MAX, etc.) on values from all the workers and then distributes the result back to every worker. Distributed training uses AllReduce operations to average all the gradients computed by all the workers in each iteration. There are two widely adopted AllReduce operations which are suitable for small-scale and large-scale scenarios respectively: Ring AllReduce(Ring) and Recursive Half-Doubling(HD). 

At the start of the Ring-allreduce\cite{cheng2019bandwidth,sergeev2018horovod} phase, each node is assigned with a rank and form a logical ring pattern according to the rank of the nodes.

For a DML occupies a whole cluster which contains $N$ GPUs, the communication phase is divided into two steps: first is scatter-reduce and second is all-gather. Scatter-reduce phase contains $N-1$ rounds and each GPU divide its local data into $N$ chunks, in round $t$, GPU $G_i$ sends its $(i-t) \bmod N$-th chunk to GPU $G_{(i+1) \bmod N}$. After $N-1$ rounds, the $(i+1) \bmod N$ chunk of GPU $G_i$  contains the local aggregated result. In all-gather phase, $G_i$ send its local aggregated in $(i+1-t) \bmod N$-th chunk to GPU $G_{d}=G_{(i+1) \bmod N}$ and after $N$-1 rounds all nodes have $N$ synchronized data chunks. Ring-allreduce follows the Leaf-wise permutation pattern. Suppose GPU $G_i$ directly connects to $l_j$, where $j = {\lfloor i/n \rfloor} = {\lfloor i*L/N \rfloor}$, $n$ is the number of ports of one Leaf and $n = N/P$. $G_d$=$G_{(i+1) \bmod N}$ directly connects to $l_{k} = l_{\lfloor ((i+1) \bmod N)/n \rfloor}$, and $\lfloor(i+1)/n\rfloor!=\lfloor i/n\rfloor$.

There exists two cases,  1) $\lfloor(i+1)/n\rfloor!=\lfloor i/n\rfloor$ and $(i+1)<N$, in this case $k={\lfloor (i+1)/n \rfloor}={\lfloor i/n \rfloor + 1}={j + 1}$. 2) $(i+1)>=N$, however $i<=N-1$, so we have $i=N-1$ and $l_k=l_0$. Suppose there exists another $G_i^*$ where $\lfloor i^*/n\rfloor!=\lfloor i/n\rfloor$ send data to $G_{d^*}=G_{(i^*+1) \bmod N}$, $G_i^*$ connects to $l_j^*=l_{\lfloor i^*/n \rfloor}$ $G_{d^*}$ connects to Leaf $l_k^*=l_{\lfloor ((i^*+1)\bmod N)/n \rfloor}$. If $k^*=k=0$, then $j=j^*=N-1$. If $k^*=k$ and $k^*!=0$, $j=j^*=k-1$. So if $k^*=k$ then we must have $j=j^*$ which contradicts the hypothesis, so Ring-allreduce follows \emph{Leaf-wise Permutation Traffic Pattern}.

For hierarchical rings or rings with multiple channels, we only need to construct an independent communication plane for each ring, where data flow in each plane follows the \emph{Leaf-wise Permutation Traffic Pattern}.

Recursive Half-Doubling (HD)\cite{2016Generalisation} is usually used when $N$ is an idempotent of two in large-scale cluster. As for HD, it contains the communication in two phases: Reduce-Scatter and AllGather.
In the Reduce-Scatter phase, it has $log_2 N$ steps. In step $t$, $i$-th rank exchanges data with a rank that is a distance $2^{t-1}(-1)^{\lfloor i/(t+1) \rfloor}$ away: Each rank sends different half of the data to its peer, and performs the reduction operation on the received data. This procedure continues recursively, halving the data communicated at each step. After the last step, each rank holds $1/N$ reduced data. 
The AllGather phase is analogous to the reduce-scatter phase. In the first step, ranks that are distance $N/2$ apart exchange their data. In the second step, ranks that are distance $N/4$ apart exchange their own data as well as the data they received in the previous step. And in step $t$, ranks that are distance $N/2^{N}$ apart exchange their data as well as the data they received in all the previous steps. After the last step, each rank hold all the reduced data. Compared to ring, HD has lower synchronization times, making it more suitable for large-scale scenarios.

In both reduce-scatter phase and AllGather phase, GPU $G_i$ send data to GPU $G_{i \oplus 2^t}$ in step $t$. Since when $t\le  \log_2n$, GPU $G_i$ and GPU $G_{i \oplus 2^t}$ directly connects to the same Leaf, so we only consider $t> \log_2n$. Take reduce-scatter phase as an example, suppose $G_i$ directly connects to $l_j$, where $l_j = b_{\lfloor i/n \rfloor}$. $G_d$=$G_{i \oplus 2^t}$  directly connects to $k={\lfloor (i \oplus 2^t)/n \rfloor}$-th Leaf. Since ${\lfloor (i \oplus 2^t)/n \rfloor} = {\lfloor (i +2^t - 2(i \land 2^t ))/n \rfloor}$. There are two cases, the first is $i \land 2^t = 0$, we have $ \lfloor i/n \rfloor \land 2^{t- \log_2n} = 0$ and ${\lfloor (i +2^t - 2(i \land 2^t ))/n \rfloor} = \lfloor(i +2^t)/n \rfloor = \lfloor i/n + 2^{t- \log_2n}\rfloor = \lfloor i/n \rfloor + 2^{t- \log_2n} = \lfloor i/n \rfloor \oplus 2^{t- \log_2n}$.  The second is  $i \land 2^t = 1$, we have $ \lfloor i/n \rfloor \land 2^{t- \log_2n} = 1$ and  ${\lfloor (i +2^t - 2(i \land 2^t ))/n \rfloor} = \lfloor(i -2^t)/n \rfloor = \lfloor i/n - 2^{t- \log_2n}\rfloor = \lfloor i/n \rfloor - 2^{t- \log_2n} = \lfloor i/n \rfloor \oplus 2^{t- \log_2n}$. All in all, $k = j \oplus 2^{t- \log_2n}$, which means some $l_j$ only communicate with one Leaf $l_{j \oplus 2^{t- \log_2n}}$ at $t$-th step, so HD also follows Leaf-wise permutation pattern.

When the number of nodes $N$ is not an idempotent of two, it is necessary for $i$-th node $G_i$, where $i=0,1,2,...,N-2^{\log_N}$ communicate with $d$-th node $G_{d}=G_{i+2^{\log_N}}$ in advance during the reduce scatter stage and the all gather phase, so that the remaining $2^{\log_N}$ nodes is an idempotent of two. Suppose GPU $G_i$ sends data to GPU $G_{d}$ through $i\%n$-th Spine, $G_i$ connects to $j$-th Leaf $l_j$, where $l_j = l_{\lfloor i/n \rfloor}$ and $n$ is the number of ports of one Leaf. $G_{
d}$  connects to $k=\lfloor (i+2^{\log_N})/n \rfloor$-th Leaf, and $\lfloor ((i+2^{\log_N})/n \rfloor != \lfloor i/n \rfloor$. Suppose there exists another $G_i^*$ send data to $G_d^* = G_{(i^*+2^{\log_N})}$ through $i^*\%n$-th Spine, where $\lfloor i/n \rfloor != \lfloor i^*/n \rfloor$, $G_d^*$ connects to $k^*=\lfloor (i^*+2^{\log_N})/n \rfloor$-th Leaf. If $i\%n = i^*\%n$, we have that $i^*-i=p*n$, where $-L<p<L$ is some non-zero integer, so $k^* = \lfloor (i+p*n+2^{\log_N})/n \rfloor=k+p$, so $k^*!=k$, so it also follows \emph{Leaf-wise Permutation Traffic Pattern}.


\paragraph{Model parallelism} With model parallelism algorithms, different part of the model are partitioned over multiple GPUs. In which each GPU is only responsible for updating the parameter weights of the neurons to which it is assigned. Different model segmentation methods may result in different communication characteristics, but in many cases \cite{fei2021efficient} the complete model will not be distributed across servers, so model parallelism usually results in intra-server traffic pattern. Even if the model has to occupy multiple servers, assigning sub-topology for each task will not lead to more serious network contention.

\paragraph{Pipeline parallelism} With pipeline parallelism, one layer of model is shared by multiple GPUs. GPU $G_{i}$ send data to GPU $G_d = G_{i+1}$ in forward phase and send data to GPU $G_{i-1}$ in backward phase. Pipeline parallelism requires send/receive between adjacent nodes. Suppose $G_i$ connects to $j$-th Leaf $l_j$, where $l_j = l_{\lfloor i/n \rfloor}$ and $n$ is the number of ports of one Leaf and $i<N$. $G_{(i + 1) }$  connects to $k=\lfloor (i + 1)/n \rfloor$-th Leaf, and $\lfloor (i + 1)/n \rfloor != \lfloor i/n \rfloor$. $G_i$ send data to $G_d$ through $i\%n$-th Spine. Suppose there exists another $G_i^*$ send data to $G_d^* = G_{(i^*+ t + 1) }$ through $i^*\%n$-th Spine, if $i\%n=i^*\%n$, then we have $-N<i^*-i=p*n<N$, so $k^*= \lfloor (i^*+ t + 1)/n \rfloor =  \lfloor (i +p*n+ t + 1)/n \rfloor=  \lfloor (i + t + 1)/n \rfloor+p$, which means $k!=k^*$, so it follows \emph{Leaf-wise Permutation Traffic Pattern}.

\paragraph{Expert parallelism} With Expert Parallelism, different groups of \emph{experts} are partitioned over multiple GPUs, and all GPUs should communicate with all other GPUs. In many cases, the communication patterns of MoE parallelism are based on pair-wise exchange algorithm. The pair-wise All2All contains $N-1$ steps. In step $t$, GPU $G_i$ sends data to GPU $G_{(i+ t + 1) \bmod N}$ through $i\%n$-th Spine. Suppose $G_i$ connects to $j$-th Leaf $l_j$, where $l_j = l_{\lfloor i/n \rfloor}$ and $n$ is the number of ports of one Leaf. $G_{(i+ t + 1) \bmod N}$  connects to $k=\lfloor ((i + t + 1) \bmod N)/n \rfloor$-th Leaf, and $\lfloor ((i + t + 1) \bmod N)/n \rfloor != \lfloor i/n \rfloor$.  Suppose there exists another $G_i^*$ send data to $G_d^* = G_{(i^*+ t + 1) \bmod N}$ through $i^*\%n$-th Spine, where $\lfloor i/n \rfloor != \lfloor i^*/n \rfloor$, $G_d^*$ connects to $k^*=\lfloor ((i^* + t + 1) \bmod N)/n \rfloor$-th Leaf. If $i\%n = i^*\%n$, we have that $i^*-i=p*n$, where $-L<p<L$ is some integer, then we have $-N<(i^*+ t + 1)-(i+ t + 1)=p*n<N$, so $k^*=\lfloor ((i^* + t + 1) \bmod N)/n \rfloor=\lfloor ((i + t + 1+p*n)\bmod N)/n $, which means $k!=k^*$, so pair-wise exchange algorithm also follows \emph{Leaf-wise Permutation Traffic Pattern}. 

For DML tasks that do not follow the \emph{Leaf-wise Permutation Traffic Pattern}, using \emph{Source Routing} can still bring performance improvements. If a Leaf Spine architecture cluster contains $L$ Leafs and $S$ Spines, in extreme cases, \emph{Source Routing} will cause at most $L$ streams to compete for bandwidth, while the ECMP strategy will cause at most $L*S$ streams to compete for bandwidth. Take Double Binary Trees based AllReduce \cite{jeaugey2019massively}as an example, it organizes ranks into two binary tree, and independently conducts a tree AllReduce of half data on two Binary tree. Each rank is a Leaf node in only one of the two binary tree. A cluster of $N$ nodes using tree communication will generate $2*N$ flows, so it is not possible to achieve exclusive links for each flow. However, through simulation, we find that in a cluster with 2048 GPUs, when using \emph{Source Routing}, the number of link contention will not exceed 3. An OCS layer can also be introduced to generated a small double-binary tree  sub-topology to eliminate network contention.

\begin{figure}[!htb]
\centering
\includegraphics[width=\linewidth]{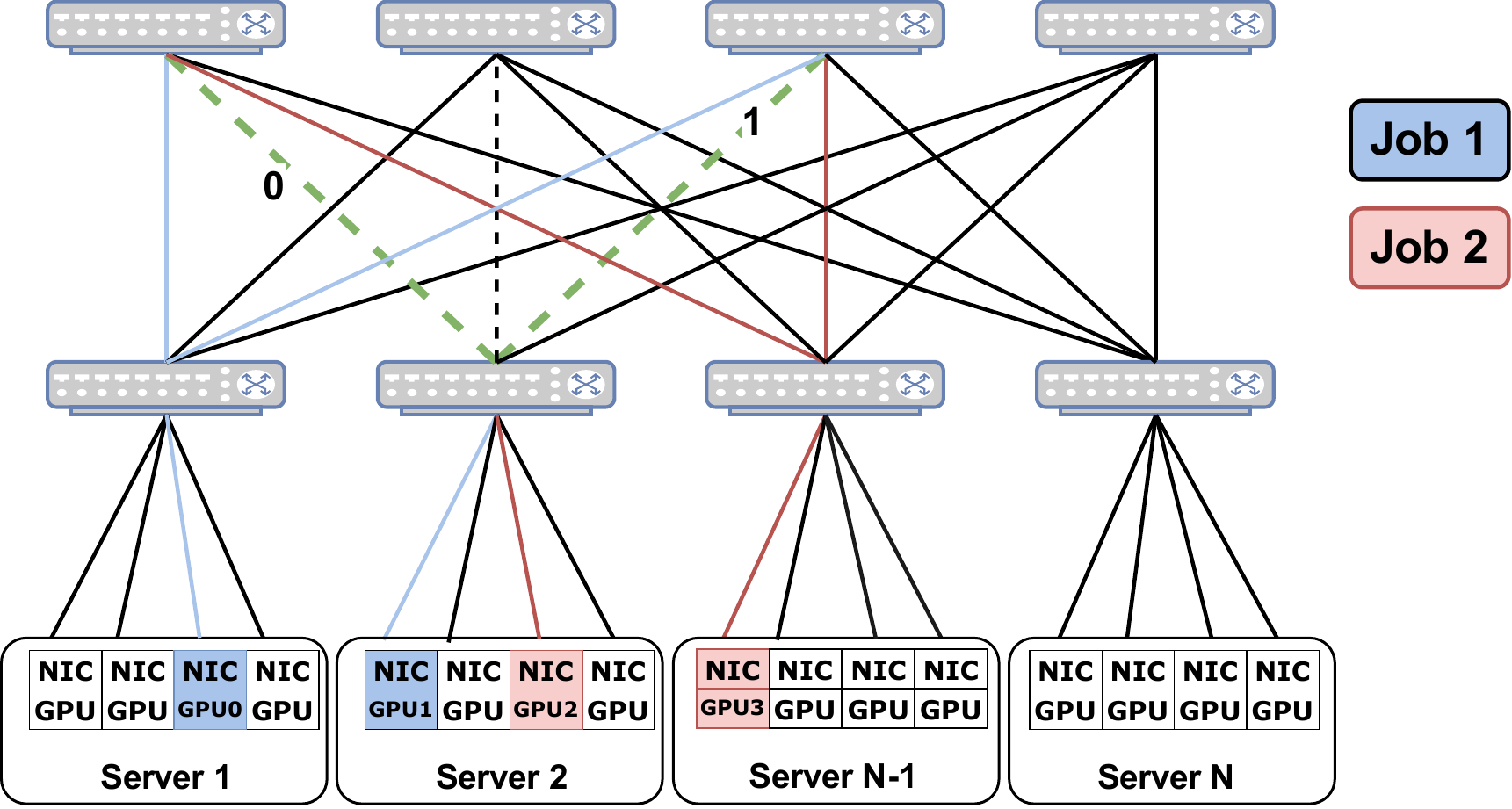}
\caption{Example of network contention when multiple tasks run in a Leaf-Spine network.}
\label{ContentionExample}
\end{figure}

When there are multiple tasks in a Leaf-Spine network, the communication pattern may not be a Leaf-wise permutation pattern. Then, we cannot guarantee contention-free. Fig. \ref{ContentionExample} shows an example, link 0 is shared when GPU0 sends flows to GPU1 and GPU3 sends flows to GPU2, link is shared when GPU2 sends flows to GPU3 and GPU1 sends flows to GPU0. 

\section{vClos Design}
\label{vclos}
\subsection{vClos: Contention-Free Multi-Task Resource Scheduling in a Leaf-Spine Network}
Single task can be deployed in a Leaf-Spine network without network contention as long as it follows \emph{Leaf-wise Permutation Traffic Pattern}. This motivates us to assign each DML task a individual virtual network which can naturally eliminate inter-task contention. To this end, we propose \emph{vClos}\footnote{A Leaf-Spine network can be viewed as a folded three-layered clos network. That is why we call our solution \emph{vClos}.}. To rigorously formulate the network resource scheduling problem in \emph{vClos}, we introduce the following definitions. Mathematical notations of \emph{vClos} is shown in Table \ref{tbl:notations}.

Let's first introduce the general process of \emph{vClos}. \emph{vClos} method runs on a cluster contain $L$ Leaf switches, $S$ Spine switches and $L*S$ GPUs. $T$ GPUs on each server support NVlink to reduce communication overhead. Each arriving $i$-th task requiring $N_i$ GPUs is assigned with a sub-topology according to the current resource usage. Meeting locality is a common condition in the industry to ensure service quality, which means for some $i$-th task with $N_i<=T$, the scheduler try to place it into a single server; some $i$-th task with $N_i>T$ is assigned with $\lceil N_i/T \rceil$ servers and a  sub-topology \emph{vClos} containing $l$ virtual Leafs and $s$ virtual Spines. The detail process of generating \emph{vClos} can be divided into two stages which is shown in Stage 2 and Stage 3. During the scheduling process, scheduler strictly followed the locality constraint. We also conducted simulation experiments on the relaxation of locality to demonstrate the necessity of meeting the locality conditions in Table. \ref{workload_oxc}.

It is worth noting that the prerequisite for successfully generating \emph{vClos} is that the number of GPUs required $N$ is a prime number. Fortunately, through analysis of the Helios\cite{hu2021characterization} dataset, we found that in the vast majority of cases, $N$ is a prime number if $N>T$. In extreme cases, we can also find an $N^{new}$, where $N^{new}$ is the smallest composite number greater than $N$, and generate a \emph{vClos} contains $N^{new}$ GPUs.

\paragraph{Stage 0: Attempt to allocate resources within one server}
Due to the locality constraint, $i$-th task with $N_i\le T$ will be placed into one server. The scheduler selects the server with the least remaining number of GPUs among all servers which are capable of placing tasks. If such a server cannot be found, the task will enter the waiting phase.

\paragraph{Stage 1: Attempt to allocate resources within one Leaf switch}
For tasks with $N_i>T$, scheduler enter into Stage 1. In order to minimize the occupation of network resources as much as possible, we first attempt to allocate tasks under the same Leaf. The advantage of this strategy is that it can reduce the occupation of the Spine port, thus reducing the fragmentation of network resources. To be more specific, scheduler considers all Leaf switches connected to more than $N_i/T$ idle servers as candidate switches. Among these candidate switches, scheduler select the one connecting least idle servers. The detailed scheduling strategy is shown in Algorithm \ref{alg:vclos}. 

        

\begin{table}[!htb]
	\centering  
	\caption{Notations used in vClos/OCS-vClos Physical Architecture}\label{tbl:notations}
	\begin{tabular}{c|l}  
		\toprule 
		$L$ & \begin{tabular}[c]{@{}l@{}}
		Num of Leafs\end{tabular}\\ 
		$S$ & \begin{tabular}[c]{@{}l@{}}
		Num of Spines\end{tabular}\\ 
		$T$ & \begin{tabular}[c]{@{}l@{}}
		Num of GPU per server\end{tabular}\\ 
		$L_n$ & \begin{tabular}[c]{@{}l@{}}
		$n-$th Leaf \end{tabular}\\ 
		$S_m$ & \begin{tabular}[c]{@{}l@{}}
		$m-$th Spine \end{tabular}\\ 
		$O_k$ & \begin{tabular}[c]{@{}l@{}}
		$k-$th OCS \end{tabular}\\ 
		$C_{n,m}$ & \begin{tabular}[c]{@{}l@{}}
		Unused links between $L_n$ and $S_m$ \end{tabular}\\ 
		$N_i$ & \begin{tabular}[c]{@{}l@{}}
		Num of GPU required by $i$-th job\end{tabular}\\ 
		$R_n$  &\begin{tabular}[c]{@{}l@{}}
		Num of remaining idle servers in $E_n$ \end{tabular}\\
		$r_n^i$  &\begin{tabular}[c]{@{}l@{}}
		Num of servers to be used in $E_n$ for $i$-th job \end{tabular}\\
		$l$  &\begin{tabular}[c]{@{}l@{}}
		Num of Virtual Leafs to be used for $i$-th job \end{tabular}\\
		$s$  &\begin{tabular}[c]{@{}l@{}}
		Num of Virtual Spines to be used for $i$-th job \end{tabular}\\
		\bottomrule
	\end{tabular}
\label{table0}        
\end{table}

\begin{algorithm}[!htb]  
	\caption{vClos Algorithm}\label{alg:vclos}
	\LinesNumbered 
	\KwIn{Required job size $N$}
	\KwOut{$gpus$, $c_{n,m}$}
	\tcc{Calculate the required size within one server} 
	$ReqSize \leftarrow min(N,T)$\; 
	\tcc{Calculated the required number of servers}
	$ReqSevNum \leftarrow \lceil N/ReqSize \rceil$\; 
	$validLeafList \leftarrow []$\;
	\tcc{Step1: Find valid Leafs with least idle servers} 
	\For{$n \leftarrow 1$ \KwTo $P$}{
		\If{$E_n.validServer(ReqSize,ReqSevNum)$}{
		    $validLeafList.append(E_n)$\;
		}
	}
	$validLeafList.sort()$\;
	\If{$len(validLeafList)\geq 1$}{
	    \tcc{Step1: Find valid servers with least idle GPUs}
	    $E_n^{*} \leftarrow validLeafList[0]$\;
	    $ gpus $ = $E_n^{*}.ChooseGpu(ReqSize,ReqSevNum)$\;
		\Return $gpus,none$\;
	}
	\tcc{Step2: Find valid vClos using ILP}
	$gpus,c_{n,m} \leftarrow FindvClos(N,Q)$\;
    \Return $gpus,c_{n,m}$\;
	
\end{algorithm}

\paragraph{Stage 2 in vClos: Attempt to generate virtual Clos}
When the scheduler is unable to place tasks under the same Leaf, it constructs \emph{vClos} through link and port reservation. In Stage 2, scheduler attempts to generate virtual Clos with $s$ virtual Spines and $l$ virtual Leafs. There are many combinations of $s$ and $l$, but as we show in Fig. \ref{fig:conflict_fig}, there is no sample path optimal solution to this problem, which means no matter how virtual clos are currently generated, there will always be task sequences that cause network resource fragmentation in the cluster in future. A direct solution to find a accessible solution is firstly choosing $l$ Leaf switches which directly connects to $N_i/T$ idle servers, then for each Leaf switch $l_j$ generating a set $s_j$ which contains all Spine switches which directly connect to $l_j$ by a unused link. If $|\cup_{s_j}|>=s$, we can choose $s$ Spine switches and then we can generate a virtual clos. Theoretically, we can try many combination of $l$ Leaf switches and $s$ Spine switches by enumerating, but it is time consuming. In Algorithm \ref{alg:vclos} we linearize this problem into an ILP problem and give a solution with an average time cost of 1 second in a cluster containing 2048 GPUs. If \emph{vClos} cannot be generated, the task will enter the waiting phase.

\section{OCS-vClos Design}
As Section \ref{vclos} shows, a simple Leaf-Spine architecture is not sufficient to eliminate network contention without incurring additional overhead. A simulation experiments where 5000 training tasks run on a cluster with 512 GPUs is conducted to analyze the impact of \emph{vClos} on the fragmentation of resources. The severity of network resource fragmentation is measured through simulation in Table \ref{tbl:fragmentation_count}, in which the arrival time of the task follows the Poisson distribution, and we modify the task arrival rate by modifying the  $\lambda$ variable of the Poisson distribution. It can be seen that when the task arrival rate is high, the network resource fragmentation introduced by \emph{vClos} may be severe.


In order to reduce the additional overhead, a layer of optical circuit switches (OCS) is introduced to rewire links between Leaf switches and Spine switches. Fig. \ref{fig:OCS-vClosExample} shows an example of \emph{OCS-vClos}. In \cite{zhao2019minimal}, integer linear programming(ILP) was used to generate re-configuring strategies for OCS, and our work drew on similar methods to rewire links to minimize network fragmentation as much as possible . 
It is worth mentioning that OCS switching requires a delay of 50ms, so occupied links can not be rewired during AI training to prevent package loss. Therefore, only idle links can be rewired to use fragmented resources.  Simulation experiment in Table 
\ref{tbl:fragmentation_count} shows that the number of fragmentation caused by network resources can be greatly reduced when using \emph{OCS-vClos}. 

Before introducing the details of \emph{OCS-vClos}, let's briefly introduce the scheduling process. To generate \emph{OCS-vClos}, the behavior of scheduler in stage0 and stage1 is similarly to those in \emph{vClos}. The scheduler places tasks with $N_i<=T$ into one server in stage 0 and try to place tasks with $N_i>T$ under one Leaf in stage 1. However, due to the flexibility brought by the OCS layer, the schedule's strategy for building virtual Clos is slightly different. In stage2, the scheduler tries to place the task under single Spine through link rewiring. If there is no valid Spine, scheduler generate \emph{OCS-vClos} using multiply Spines through rewiring links in stage 3.

\begin{table}[!tbp]
\centering
\caption{Fragmentation Count with Causes}
\label{tbl:fragmentation_count}
\begin{tabular}{ccccc}
\toprule
\multirow{2}{*}{$\lambda(s)$} & \multicolumn{2}{c}{vClos} & \multicolumn{2}{c}{OCS-vClos} \\
                         & GPU        & Network      & GPU          & Network        \\
\midrule
100                        & 1311       & 264          & 1180         & 172            \\
110                        & 988        & 250          & 1047         & 161            \\
120                        & 878        & 218          & 920          & 129            \\
130                        & 726        & 145          & 705          & 98             \\
\bottomrule
\end{tabular}
\end{table}


\subsection{Stage 0 and Stage 1 in OCS-vClos}
When dealing with tasks with $N_i<=T$, the scheduler enter into Stage 0 and tries to place the task into one server with least idle GPUs. For tasks with $N_i>T$, in Stage 1, the scheduler tries to select the Leaf switch with the least number of idle servers connected among all leaves that can place tasks. If there is no valid Leaf, scheduler tries to place the task under one Spine in Stage 2.

\subsection{Stage 2 in OCS-vClos: Attempt to generate virtual Clos within one Spine}
In Stage 2 \emph{OCS-vClos} tries to generate virtual Clos within one Spine. Benefit from the flexibility of OCS reconfiguration, \emph{OCS-vClos} can generate virtual Clos within one Spine. Stage 2 can be divided into 2 steps, firstly \emph{OCS-vClos} choose the Spine with least but enough ports. When the Spine $S_m$ is chosen, OCS $O_k$ can be used to connect to some server which belongs to Leaf $L_n$ and this Spine $S_m$ as long as there are idle ports connecting to $L_n$ and $S_m$. Secondly, \emph{OCS-vClos} choose the servers by rewiring links without affecting current tasks.
It is worth mentioning that if a task only uses two Leafs, a special \emph{OCS-vClos} as shown in Fig. \ref{fig:hash_collison} can be generated directly through OCS reconfiguration without using Spines. This can reduce the usage of Spine ports and further reduce network resource fragmentation.

\subsection{Stage 3 in OCS-vClos: Attempt to generate virtual Clos}
If virtual Clos can not be generated within one Spine, \emph{OCS-vClos} tries to generate \emph{vClos} with $l$ Leafs and $s$ Spines. Similarly in \emph{vClos}, we can try to enumerate all possible combinations of p, q and OCS configuration, but it is time consuming.  In Algorithm \ref{alg:ocs_vclos} we linearize this problem into an ILP problem and give a solution with an average time cost of 2 second in a cluster containing 2048 GPUs.

\begin{algorithm}[!htb] 
	\caption{OCS-vClos Algorithm}\label{alg:ocs_vclos}
	\LinesNumbered 
	\KwIn{Required job size $N_i$}
	\KwOut{$gpus$, $c_{n,m}^k$}
	\tcc{Calculate the required size within one server} 
	$ServerSize \leftarrow min(N_i,T)$\;
	\tcc{Calculated the required number of servers}
	$ReqSevNum \leftarrow \lceil N_i/ServerSize \rceil$\;
	\tcc{Step1: Find valid Leafs with least idle servers} 
	$validLeafList \leftarrow []$\;
	\For{$n \leftarrow 1$ \KwTo $L$}{
		\If{$L_n.validServer(ServerSize,ReqSevNum)$}{
		    $validLeafList.append(L_n)$\;
		}
	}
	$validLeafList.sort()$\;
	\If{$len(validLeafList)\geq 1$}{
	    \tcc{Step1: Find servers Leafs with least idle GPUs} 
	    $L_n^{*} \leftarrow validLeafList[0]$\;
	    $ gpus $ = $L_n^{*}.ChooseGpu(ServerSize,ReqSevNum)$\;
		\Return $gpus,none$\;
	}
	\tcc{Step2: Find vClos within one Spine} 
	$validSpineList \leftarrow []$\;
	\For{$m \leftarrow 1$ \KwTo $S$}{
		\If{$S_m.freePort>N_i$}{
		    $validSpineList.append(S_m)$\;
		}
	}
	$validSpineList.sort()$\;
	\tcc{Step3: Find valid vClos using ILP} 
	\If{$len(validSpineList)\geq 1$}{
	    $S_m^{*} \leftarrow validSpineList[0]$\;
	    $ gpus, c_{n,m}^k $ = $S_m^{*}.ChooseGpu(T,ReqSevNum)$\;
		\Return $gpus,c_{n,m}^k$.
	}
	$gpus, c_{n,m}^k \leftarrow OCS-FindClos(N_i,S)$\;
	\Return $gpus, c_{n,m}^k$\;
\end{algorithm}

\section{Implementation}
\label{impla}

\subsection{Simulation Setup}
\textbf{Traning hardware}. Our experiment contains both testbed experiment and simulation experiment. The testbed experiment is conducted on 8 Huawei FusionServer G5500 servers. Each server contains 2 Intel Xeon processors and 64GB memory. In order to avoid potential PCIe conflicts, we placed 4 NVIDIAS Tesla V100 GPUs on each server, with each GPU bound to a Mellanox-5 network card that can provide 100gbps of bandwidth. We use four 32 port Huawei CE8850 switches to form a Leaf Spine network, and we add a MEMS-OCS between the Leaf and Spine layers. 

\textbf{Generation of Task Datasets in Testbed Experiments}. Our test dataset includes 100 tasks including data parallel tasks (VGG16, ResNet50, ResNet101), MoE and DLRM, where the batch size of different DML is shown in Table \ref{tbl:testbed_batchsize}. The communication algorithm is randomly selected among Ring, hierarchical Ring, and HD. Communication is based on PyTorch's built-in communication library DDP. The number of GPUs required for the task is randomly selected from 2, 4, 8, and 16. FIFO is chosen as the default job scheduler.



\begin{table}[!tbp]
	\centering 
	\caption{mini-BatchSize set in testbed experiment}\label{tbl:testbed_batchsize}
	\setlength{\tabcolsep}{1.0mm}
	\begin{tabular}{c|ccccc}
	\toprule
	    $TaskType$ &VGG16&ResNet50/101&BERT&MoE&DLRM \\
	    \midrule
		$BatchSize$& 16,32 & 32,64  & 4,8 &8,16 & 256,512 \\  
		\bottomrule 
	\end{tabular}
\end{table}

\textbf{Implementation of Communication Algorithm}. For data parallel tasks, we used Horovod 0.19.2 and a modified version of NCCL 2.7.8 for distributed training. For some commonly used communication algorithms in the industry, such as hierarchical ring and HD algorithms, NCCL does not have native support. Therefore, we have added a custom communication algorithm design layer on top of the NCCL 2.7.8 communication library to support hierarchical ring and HD algorithms. For tasks that require All2All communication such as DLRM and MoE, Torch 1.8.1 is used to run and NCCL and pytorch ddp is used for communication. 

\textbf{Implementation of Routing Scheme}. We tested four different routing schemes, namely \emph{OCS-vClos}, \emph{Source Routing}, ECMP, and the ECMP strategy of adding additional 50\% ports. Because we don't have enough switches to test large-scale networking, we use VRF to flexibly virtualize a switch into multiple switches. 
In \emph{OCS-vClos} and \emph{Source Routing} experiment, we configure the ACL command on the incoming port of the switch to realize \emph{Source Routing}. Four independent Routing table are configured in the Spine switch through VRF, and one Spine switch is virtualized into four logical Spine switches. Inter-task contention may occur when using \emph{Source Routing}.

The scheduler will reconfigure the network based on the current task requirements and network status. OCS' port reconfiguration in our experiment relays on a Telnet-based protocol. A trivial procedure has been developed, allowing for the declarative alteration of OCS links while preserving the integrity of connections that are presently in use, ensuring uninterrupted connectivity and re-configurability throughout the network.

Due to the stochastic nature of the hashing algorithms used in ECMP, contention is inevitable, especially when communication pattern are arbitrary and cannot be determined a priori. As the specific hashing algorithm remains undisclosed, we undertake a series of experiments as referred in Section \ref{hash-collision-chosen} to determine an optimum combination of hashing mode and factors. In these experiments, we randomly select communication peers and analyze the distribution of data flows among the interconnected links that possess equal costs. The 5-tuple \textsf{(src-ip,dst-ip,src-port,dst-port,port number)} is chosen as the hash factor and hash mode is set to 5. 

\subsection{Performance Analysis in Testbed Experiment}
\begin{figure}[!htbp]
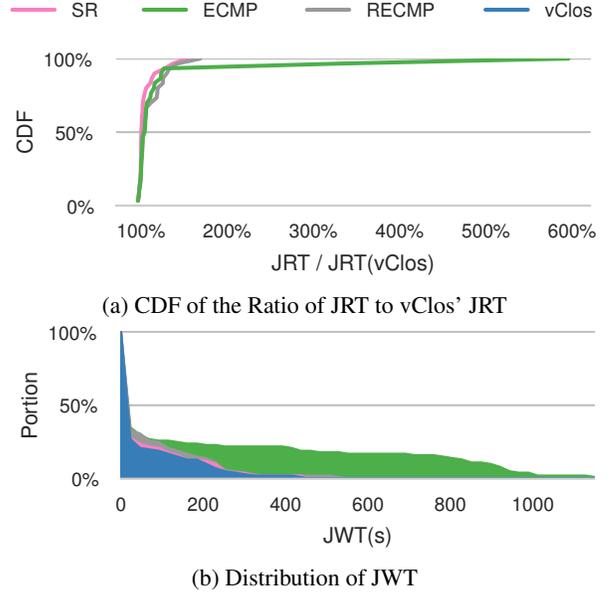

\centering
\begingroup%
\makeatletter%
\begin{pgfpicture}%
\pgfpathrectangle{\pgfpointorigin}{\pgfqpoint{3.154460in}{0.385000in}}%
\pgfusepath{use as bounding box, clip}%
\begin{pgfscope}%
\pgfsetbuttcap%
\pgfsetmiterjoin%
\definecolor{currentfill}{rgb}{1.000000,1.000000,1.000000}%
\pgfsetfillcolor{currentfill}%
\pgfsetlinewidth{0.000000pt}%
\definecolor{currentstroke}{rgb}{1.000000,1.000000,1.000000}%
\pgfsetstrokecolor{currentstroke}%
\pgfsetdash{}{0pt}%
\pgfpathmoveto{\pgfqpoint{0.000000in}{0.000000in}}%
\pgfpathlineto{\pgfqpoint{3.154460in}{0.000000in}}%
\pgfpathlineto{\pgfqpoint{3.154460in}{0.385000in}}%
\pgfpathlineto{\pgfqpoint{0.000000in}{0.385000in}}%
\pgfpathlineto{\pgfqpoint{0.000000in}{0.000000in}}%
\pgfpathclose%
\pgfusepath{fill}%
\end{pgfscope}%
\begin{pgfscope}%
\pgfsetroundcap%
\pgfsetroundjoin%
\pgfsetlinewidth{1.505625pt}%
\definecolor{currentstroke}{rgb}{0.968627,0.505882,0.749020}%
\pgfsetstrokecolor{currentstroke}%
\pgfsetdash{}{0pt}%
\pgfpathmoveto{\pgfqpoint{0.044444in}{0.202662in}}%
\pgfpathlineto{\pgfqpoint{0.155556in}{0.202662in}}%
\pgfpathlineto{\pgfqpoint{0.266667in}{0.202662in}}%
\pgfusepath{stroke}%
\end{pgfscope}%
\begin{pgfscope}%
\definecolor{textcolor}{rgb}{0.150000,0.150000,0.150000}%
\pgfsetstrokecolor{textcolor}%
\pgfsetfillcolor{textcolor}%
\pgftext[x=0.355556in,y=0.163773in,left,base]{\color{textcolor}\sffamily\fontsize{8.000000}{9.600000}\selectfont SR}%
\end{pgfscope}%
\begin{pgfscope}%
\pgfsetroundcap%
\pgfsetroundjoin%
\pgfsetlinewidth{1.505625pt}%
\definecolor{currentstroke}{rgb}{0.301961,0.686275,0.290196}%
\pgfsetstrokecolor{currentstroke}%
\pgfsetdash{}{0pt}%
\pgfpathmoveto{\pgfqpoint{0.732129in}{0.202662in}}%
\pgfpathlineto{\pgfqpoint{0.843240in}{0.202662in}}%
\pgfpathlineto{\pgfqpoint{0.954351in}{0.202662in}}%
\pgfusepath{stroke}%
\end{pgfscope}%
\begin{pgfscope}%
\definecolor{textcolor}{rgb}{0.150000,0.150000,0.150000}%
\pgfsetstrokecolor{textcolor}%
\pgfsetfillcolor{textcolor}%
\pgftext[x=1.043240in,y=0.163773in,left,base]{\color{textcolor}\sffamily\fontsize{8.000000}{9.600000}\selectfont ECMP}%
\end{pgfscope}%
\begin{pgfscope}%
\pgfsetroundcap%
\pgfsetroundjoin%
\pgfsetlinewidth{1.505625pt}%
\definecolor{currentstroke}{rgb}{0.600000,0.600000,0.600000}%
\pgfsetstrokecolor{currentstroke}%
\pgfsetdash{}{0pt}%
\pgfpathmoveto{\pgfqpoint{1.586480in}{0.202662in}}%
\pgfpathlineto{\pgfqpoint{1.697591in}{0.202662in}}%
\pgfpathlineto{\pgfqpoint{1.808702in}{0.202662in}}%
\pgfusepath{stroke}%
\end{pgfscope}%
\begin{pgfscope}%
\definecolor{textcolor}{rgb}{0.150000,0.150000,0.150000}%
\pgfsetstrokecolor{textcolor}%
\pgfsetfillcolor{textcolor}%
\pgftext[x=1.897591in,y=0.163773in,left,base]{\color{textcolor}\sffamily\fontsize{8.000000}{9.600000}\selectfont RECMP}%
\end{pgfscope}%
\begin{pgfscope}%
\pgfsetroundcap%
\pgfsetroundjoin%
\pgfsetlinewidth{1.505625pt}%
\definecolor{currentstroke}{rgb}{0.215686,0.494118,0.721569}%
\pgfsetstrokecolor{currentstroke}%
\pgfsetdash{}{0pt}%
\pgfpathmoveto{\pgfqpoint{2.521072in}{0.202662in}}%
\pgfpathlineto{\pgfqpoint{2.632183in}{0.202662in}}%
\pgfpathlineto{\pgfqpoint{2.743294in}{0.202662in}}%
\pgfusepath{stroke}%
\end{pgfscope}%
\begin{pgfscope}%
\definecolor{textcolor}{rgb}{0.150000,0.150000,0.150000}%
\pgfsetstrokecolor{textcolor}%
\pgfsetfillcolor{textcolor}%
\pgftext[x=2.832183in,y=0.163773in,left,base]{\color{textcolor}\sffamily\fontsize{8.000000}{9.600000}\selectfont vClos}%
\end{pgfscope}%
\end{pgfpicture}%
\makeatother%
\endgroup%
\begin{subfigure}[b]{\linewidth}
     \centering
     \input{images/testbed_result/JRT_ratio.pgf}
     \caption{CDF of the Ratio of JRT to vClos' JRT}
 \end{subfigure}
 
 \begin{subfigure}[b]{\linewidth}
     \centering
     \input{images/testbed_result/JWT_distribution.pgf}
     \caption{Distribution of JWT}
 \end{subfigure}
 
 \caption{Tasks Performance for DML requires more than 8 GPUs.}\label{JRT_ratio}
\end{figure}

In the testbed experiment, we additionally tested the performance of ECMP when there is an increase of 50\% links between the leaves and Spines. The results are shown in Table \ref{tbl:testbed_result} and Fig. \ref{JRT_ratio}, and it can be seen that the performance of \emph{OCS-vClos}, \emph{vClos}, and \emph{Source Routing} is better than ECMP. Even if there are more redundant ports when using ECMP-redundance (rECMP) strategy, network contention still cannot be avoided. 

\begin{table}[!tbp]
	\centering 
	\caption{Performance on Testbed Environment(s)}\label{tbl:testbed_result}
	\begin{tabular}{c|cccc}
	    \toprule  
	    Strategy&ECMP&Redundance&SR&\textbf{vClos}\\
	    \midrule  
		$Avg. JRT$& 95.12 & 86.23 & 76.92 & \textbf{74.72}\\  
		$Avg. JWT$& 72.64 & 60.73 & 50.19 & \textbf{45.56}\\   
		\bottomrule 
	\end{tabular}
\end{table}

What is more, we found that when using ECMP, $Avg. JRT$ for tasks requiring 8 GPUs increase by 8.1\%, while $Avg. JRT$ of tasks requiring 16 GPUs increased by 17.9\%, which may be because AI communication is an all or nothing task, and as long as there exists two flow compete bandwidth, the training time will be affected. This phenomenon suggests that our large tasks may be more susceptible to network contention.

\section{Large-Scale Simulation Result}

\subsection{Simulation Experiment Accuracy Adjustment}
Since fine-grained simulation runs very slowly facing plentiful flows generated by large-scale DML jobs, we develop a coarse-grained simulator RapidNetSim\cite{RapidNetSim} for large-scale simulation. RapidNetSim is a flow-level network simulation framework that enables fast
large-scale simulation with minimal accuracy loss. Compared to traditional data centers, GPU clusters typically enable policies such as PFC, while no deadlocks occur in \emph{vClos} and \emph{OCS-vClos}, so the network environment is relatively simple, and it is possible to use RapidNetSim in large-scale GPU cluster.

Fig. \ref{fig:contention_impact} shows the proportion of different types of tasks slowed down by network communication, so the uncoverable communication proportion $\alpha$ is essential for a reliable simulation results. We set $\alpha$ value for different tasks in simulation according to the testbed result.

\begin{figure}[!t]
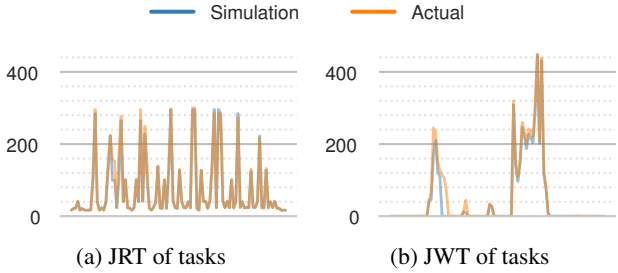

    \centering
\begingroup%
\makeatletter%
\begin{pgfpicture}%
\pgfpathrectangle{\pgfpointorigin}{\pgfqpoint{2.480000in}{0.385000in}}%
\pgfusepath{use as bounding box, clip}%
\begin{pgfscope}%
\pgfsetbuttcap%
\pgfsetmiterjoin%
\definecolor{currentfill}{rgb}{1.000000,1.000000,1.000000}%
\pgfsetfillcolor{currentfill}%
\pgfsetlinewidth{0.000000pt}%
\definecolor{currentstroke}{rgb}{1.000000,1.000000,1.000000}%
\pgfsetstrokecolor{currentstroke}%
\pgfsetdash{}{0pt}%
\pgfpathmoveto{\pgfqpoint{0.000000in}{0.000000in}}%
\pgfpathlineto{\pgfqpoint{2.480000in}{0.000000in}}%
\pgfpathlineto{\pgfqpoint{2.480000in}{0.385000in}}%
\pgfpathlineto{\pgfqpoint{0.000000in}{0.385000in}}%
\pgfpathlineto{\pgfqpoint{0.000000in}{0.000000in}}%
\pgfpathclose%
\pgfusepath{fill}%
\end{pgfscope}%
\begin{pgfscope}%
\pgfsetroundcap%
\pgfsetroundjoin%
\pgfsetlinewidth{1.505625pt}%
\definecolor{currentstroke}{rgb}{0.215686,0.494118,0.721569}%
\pgfsetstrokecolor{currentstroke}%
\pgfsetdash{}{0pt}%
\pgfpathmoveto{\pgfqpoint{0.403987in}{0.202662in}}%
\pgfpathlineto{\pgfqpoint{0.515098in}{0.202662in}}%
\pgfpathlineto{\pgfqpoint{0.626209in}{0.202662in}}%
\pgfusepath{stroke}%
\end{pgfscope}%
\begin{pgfscope}%
\definecolor{textcolor}{rgb}{0.150000,0.150000,0.150000}%
\pgfsetstrokecolor{textcolor}%
\pgfsetfillcolor{textcolor}%
\pgftext[x=0.715098in,y=0.163773in,left,base]{\color{textcolor}\sffamily\fontsize{8.000000}{9.600000}\selectfont Simulation}%
\end{pgfscope}%
\begin{pgfscope}%
\pgfsetroundcap%
\pgfsetroundjoin%
\pgfsetlinewidth{1.505625pt}%
\definecolor{currentstroke}{rgb}{1.000000,0.498039,0.000000}%
\pgfsetstrokecolor{currentstroke}%
\pgfsetdash{}{0pt}%
\pgfpathmoveto{\pgfqpoint{1.456092in}{0.202662in}}%
\pgfpathlineto{\pgfqpoint{1.567203in}{0.202662in}}%
\pgfpathlineto{\pgfqpoint{1.678314in}{0.202662in}}%
\pgfusepath{stroke}%
\end{pgfscope}%
\begin{pgfscope}%
\definecolor{textcolor}{rgb}{0.150000,0.150000,0.150000}%
\pgfsetstrokecolor{textcolor}%
\pgfsetfillcolor{textcolor}%
\pgftext[x=1.767203in,y=0.163773in,left,base]{\color{textcolor}\sffamily\fontsize{8.000000}{9.600000}\selectfont Actual}%
\end{pgfscope}%
\end{pgfpicture}%
\makeatother%
\endgroup%
    
    \begin{subfigure}[b]{0.49\linewidth}
     \centering
     \input{images/sim_result/JRT_rapid.pgf}
     \caption{JRT of tasks}
    \end{subfigure}
    \begin{subfigure}[b]{0.49\linewidth}
     \centering
     \input{images/sim_result/JWT_rapid.pgf}
     \caption{JWT of tasks}
    \end{subfigure}
   \caption{Consistency between Simulation and Actual Results, SR Strategy}\label{Adjustment} 
\end{figure}

Fig. \ref{Adjustment} shows the distribution of $JRT$ and $JWT$ on RapidnetSim for 100 jobs tested in testbed experiment, the results of the two experiments are basically consistent.

\subsection{Simulation Setup}
The simulation experiment is conducted on clusters of 512 GPU \textsc{Cluster512} and 2048 GPU \textsc{Cluster2048}  with a coarse grained simulator RapidNetSim. The dataset which contains 5000 DML tasks is based on Helios\cite{hu2021characterization}. Although the Helios dataset contains task arrival time information, the cluster size we tested is different from that in\cite{hu2021characterization}, so in our simulation the arrive time of each tasks follows a  Poisson distribution with $\lambda=120s$. FIFO is chosen as the default job scheduler.
We have also tried to conduct simulations based on the fine-grained simulator OMNET, but the simulation speed is too slow. Therefore, we chose RapidNetSim for large-scale networking simulation experiments. 

\begin{figure*}[!t]
    \centering
\begingroup%
\makeatletter%
\begin{pgfpicture}%
\pgfpathrectangle{\pgfpointorigin}{\pgfqpoint{5.413021in}{0.385000in}}%
\pgfusepath{use as bounding box, clip}%
\begin{pgfscope}%
\pgfsetbuttcap%
\pgfsetmiterjoin%
\definecolor{currentfill}{rgb}{1.000000,1.000000,1.000000}%
\pgfsetfillcolor{currentfill}%
\pgfsetlinewidth{0.000000pt}%
\definecolor{currentstroke}{rgb}{1.000000,1.000000,1.000000}%
\pgfsetstrokecolor{currentstroke}%
\pgfsetdash{}{0pt}%
\pgfpathmoveto{\pgfqpoint{0.000000in}{0.000000in}}%
\pgfpathlineto{\pgfqpoint{5.413021in}{0.000000in}}%
\pgfpathlineto{\pgfqpoint{5.413021in}{0.385000in}}%
\pgfpathlineto{\pgfqpoint{0.000000in}{0.385000in}}%
\pgfpathlineto{\pgfqpoint{0.000000in}{0.000000in}}%
\pgfpathclose%
\pgfusepath{fill}%
\end{pgfscope}%
\begin{pgfscope}%
\pgfsetroundcap%
\pgfsetroundjoin%
\pgfsetlinewidth{1.505625pt}%
\definecolor{currentstroke}{rgb}{0.215686,0.494118,0.721569}%
\pgfsetstrokecolor{currentstroke}%
\pgfsetdash{}{0pt}%
\pgfpathmoveto{\pgfqpoint{0.044444in}{0.202662in}}%
\pgfpathlineto{\pgfqpoint{0.155556in}{0.202662in}}%
\pgfpathlineto{\pgfqpoint{0.266667in}{0.202662in}}%
\pgfusepath{stroke}%
\end{pgfscope}%
\begin{pgfscope}%
\definecolor{textcolor}{rgb}{0.150000,0.150000,0.150000}%
\pgfsetstrokecolor{textcolor}%
\pgfsetfillcolor{textcolor}%
\pgftext[x=0.355556in,y=0.163773in,left,base]{\color{textcolor}\sffamily\fontsize{8.000000}{9.600000}\selectfont vClos}%
\end{pgfscope}%
\begin{pgfscope}%
\pgfsetroundcap%
\pgfsetroundjoin%
\pgfsetlinewidth{1.505625pt}%
\definecolor{currentstroke}{rgb}{1.000000,0.498039,0.000000}%
\pgfsetstrokecolor{currentstroke}%
\pgfsetdash{}{0pt}%
\pgfpathmoveto{\pgfqpoint{0.855610in}{0.202662in}}%
\pgfpathlineto{\pgfqpoint{0.966721in}{0.202662in}}%
\pgfpathlineto{\pgfqpoint{1.077832in}{0.202662in}}%
\pgfusepath{stroke}%
\end{pgfscope}%
\begin{pgfscope}%
\definecolor{textcolor}{rgb}{0.150000,0.150000,0.150000}%
\pgfsetstrokecolor{textcolor}%
\pgfsetfillcolor{textcolor}%
\pgftext[x=1.166721in,y=0.163773in,left,base]{\color{textcolor}\sffamily\fontsize{8.000000}{9.600000}\selectfont OCS-vClos}%
\end{pgfscope}%
\begin{pgfscope}%
\pgfsetroundcap%
\pgfsetroundjoin%
\pgfsetlinewidth{1.505625pt}%
\definecolor{currentstroke}{rgb}{0.301961,0.686275,0.290196}%
\pgfsetstrokecolor{currentstroke}%
\pgfsetdash{}{0pt}%
\pgfpathmoveto{\pgfqpoint{1.944553in}{0.202662in}}%
\pgfpathlineto{\pgfqpoint{2.055664in}{0.202662in}}%
\pgfpathlineto{\pgfqpoint{2.166775in}{0.202662in}}%
\pgfusepath{stroke}%
\end{pgfscope}%
\begin{pgfscope}%
\definecolor{textcolor}{rgb}{0.150000,0.150000,0.150000}%
\pgfsetstrokecolor{textcolor}%
\pgfsetfillcolor{textcolor}%
\pgftext[x=2.255664in,y=0.163773in,left,base]{\color{textcolor}\sffamily\fontsize{8.000000}{9.600000}\selectfont ECMP}%
\end{pgfscope}%
\begin{pgfscope}%
\pgfsetroundcap%
\pgfsetroundjoin%
\pgfsetlinewidth{1.505625pt}%
\definecolor{currentstroke}{rgb}{0.968627,0.505882,0.749020}%
\pgfsetstrokecolor{currentstroke}%
\pgfsetdash{}{0pt}%
\pgfpathmoveto{\pgfqpoint{2.798904in}{0.202662in}}%
\pgfpathlineto{\pgfqpoint{2.910015in}{0.202662in}}%
\pgfpathlineto{\pgfqpoint{3.021126in}{0.202662in}}%
\pgfusepath{stroke}%
\end{pgfscope}%
\begin{pgfscope}%
\definecolor{textcolor}{rgb}{0.150000,0.150000,0.150000}%
\pgfsetstrokecolor{textcolor}%
\pgfsetfillcolor{textcolor}%
\pgftext[x=3.110015in,y=0.163773in,left,base]{\color{textcolor}\sffamily\fontsize{8.000000}{9.600000}\selectfont SR}%
\end{pgfscope}%
\begin{pgfscope}%
\pgfsetroundcap%
\pgfsetroundjoin%
\pgfsetlinewidth{1.505625pt}%
\definecolor{currentstroke}{rgb}{0.650980,0.337255,0.156863}%
\pgfsetstrokecolor{currentstroke}%
\pgfsetdash{}{0pt}%
\pgfpathmoveto{\pgfqpoint{3.486588in}{0.202662in}}%
\pgfpathlineto{\pgfqpoint{3.597699in}{0.202662in}}%
\pgfpathlineto{\pgfqpoint{3.708811in}{0.202662in}}%
\pgfusepath{stroke}%
\end{pgfscope}%
\begin{pgfscope}%
\definecolor{textcolor}{rgb}{0.150000,0.150000,0.150000}%
\pgfsetstrokecolor{textcolor}%
\pgfsetfillcolor{textcolor}%
\pgftext[x=3.797699in,y=0.163773in,left,base]{\color{textcolor}\sffamily\fontsize{8.000000}{9.600000}\selectfont Best}%
\end{pgfscope}%
\begin{pgfscope}%
\pgfsetroundcap%
\pgfsetroundjoin%
\pgfsetlinewidth{1.505625pt}%
\definecolor{currentstroke}{rgb}{0.596078,0.305882,0.639216}%
\pgfsetstrokecolor{currentstroke}%
\pgfsetdash{}{0pt}%
\pgfpathmoveto{\pgfqpoint{4.242252in}{0.202662in}}%
\pgfpathlineto{\pgfqpoint{4.353363in}{0.202662in}}%
\pgfpathlineto{\pgfqpoint{4.464475in}{0.202662in}}%
\pgfusepath{stroke}%
\end{pgfscope}%
\begin{pgfscope}%
\definecolor{textcolor}{rgb}{0.150000,0.150000,0.150000}%
\pgfsetstrokecolor{textcolor}%
\pgfsetfillcolor{textcolor}%
\pgftext[x=4.553363in,y=0.163773in,left,base]{\color{textcolor}\sffamily\fontsize{8.000000}{9.600000}\selectfont Balanced ECMP}%
\end{pgfscope}%
\end{pgfpicture}%
\makeatother%
\endgroup%
    
    \begin{subfigure}[b]{0.49\linewidth}
     \centering
     \input{images/sim_result/JRT_120.pgf}
     
     {\small \textsf{Long Tail effect caused by hash-collision affects \emph{Stability}.}}
     \caption{CDF and Distribution of JRT}\label{fig:JRT}
    \end{subfigure}
    \begin{subfigure}[b]{0.49\linewidth}
     \centering
     \input{images/sim_result/JWT_120.pgf}
     
     {\small \textsf{Long Tail effect is less observed in \emph{OCS-vClos} or \emph{vClos}.}}
     \caption{CDF and Distribution of JWT}\label{fig:JWT}
    \end{subfigure}
    
    \begin{subfigure}[b]{0.49\linewidth}
     \centering
     \input{images/sim_result/QL_120.pgf}
     
     {\small \textsf{Hash-collision may cause a significant increase in waiting queue depth, affecting JWT and user experience.}}
     \caption{Queue Depth}\label{fig:queue}
    \end{subfigure}
    \begin{subfigure}[b]{0.49\linewidth}
     \centering
     \input{images/sim_result/stability_120.pgf}
     
     {\small \textsf{Eliminating network contention can effectively improve task \emph{Stability}, especially for DML with more GPUs.}}
     \caption{Stability}\label{fig:stab_512}
    \end{subfigure}
    
    \caption{Performance of 5000 tasks on \textsc{Cluster512}}\label{Sim_120_512}
\end{figure*}

\begin{figure*}[!t] 
    \centering
\begingroup%
\makeatletter%
\begin{pgfpicture}%
\pgfpathrectangle{\pgfpointorigin}{\pgfqpoint{5.413021in}{0.385000in}}%
\pgfusepath{use as bounding box, clip}%
\begin{pgfscope}%
\pgfsetbuttcap%
\pgfsetmiterjoin%
\definecolor{currentfill}{rgb}{1.000000,1.000000,1.000000}%
\pgfsetfillcolor{currentfill}%
\pgfsetlinewidth{0.000000pt}%
\definecolor{currentstroke}{rgb}{1.000000,1.000000,1.000000}%
\pgfsetstrokecolor{currentstroke}%
\pgfsetdash{}{0pt}%
\pgfpathmoveto{\pgfqpoint{0.000000in}{0.000000in}}%
\pgfpathlineto{\pgfqpoint{5.413021in}{0.000000in}}%
\pgfpathlineto{\pgfqpoint{5.413021in}{0.385000in}}%
\pgfpathlineto{\pgfqpoint{0.000000in}{0.385000in}}%
\pgfpathlineto{\pgfqpoint{0.000000in}{0.000000in}}%
\pgfpathclose%
\pgfusepath{fill}%
\end{pgfscope}%
\begin{pgfscope}%
\pgfsetroundcap%
\pgfsetroundjoin%
\pgfsetlinewidth{1.505625pt}%
\definecolor{currentstroke}{rgb}{0.215686,0.494118,0.721569}%
\pgfsetstrokecolor{currentstroke}%
\pgfsetdash{}{0pt}%
\pgfpathmoveto{\pgfqpoint{0.044444in}{0.202662in}}%
\pgfpathlineto{\pgfqpoint{0.155556in}{0.202662in}}%
\pgfpathlineto{\pgfqpoint{0.266667in}{0.202662in}}%
\pgfusepath{stroke}%
\end{pgfscope}%
\begin{pgfscope}%
\definecolor{textcolor}{rgb}{0.150000,0.150000,0.150000}%
\pgfsetstrokecolor{textcolor}%
\pgfsetfillcolor{textcolor}%
\pgftext[x=0.355556in,y=0.163773in,left,base]{\color{textcolor}\sffamily\fontsize{8.000000}{9.600000}\selectfont vClos}%
\end{pgfscope}%
\begin{pgfscope}%
\pgfsetroundcap%
\pgfsetroundjoin%
\pgfsetlinewidth{1.505625pt}%
\definecolor{currentstroke}{rgb}{1.000000,0.498039,0.000000}%
\pgfsetstrokecolor{currentstroke}%
\pgfsetdash{}{0pt}%
\pgfpathmoveto{\pgfqpoint{0.855610in}{0.202662in}}%
\pgfpathlineto{\pgfqpoint{0.966721in}{0.202662in}}%
\pgfpathlineto{\pgfqpoint{1.077832in}{0.202662in}}%
\pgfusepath{stroke}%
\end{pgfscope}%
\begin{pgfscope}%
\definecolor{textcolor}{rgb}{0.150000,0.150000,0.150000}%
\pgfsetstrokecolor{textcolor}%
\pgfsetfillcolor{textcolor}%
\pgftext[x=1.166721in,y=0.163773in,left,base]{\color{textcolor}\sffamily\fontsize{8.000000}{9.600000}\selectfont OCS-vClos}%
\end{pgfscope}%
\begin{pgfscope}%
\pgfsetroundcap%
\pgfsetroundjoin%
\pgfsetlinewidth{1.505625pt}%
\definecolor{currentstroke}{rgb}{0.301961,0.686275,0.290196}%
\pgfsetstrokecolor{currentstroke}%
\pgfsetdash{}{0pt}%
\pgfpathmoveto{\pgfqpoint{1.944553in}{0.202662in}}%
\pgfpathlineto{\pgfqpoint{2.055664in}{0.202662in}}%
\pgfpathlineto{\pgfqpoint{2.166775in}{0.202662in}}%
\pgfusepath{stroke}%
\end{pgfscope}%
\begin{pgfscope}%
\definecolor{textcolor}{rgb}{0.150000,0.150000,0.150000}%
\pgfsetstrokecolor{textcolor}%
\pgfsetfillcolor{textcolor}%
\pgftext[x=2.255664in,y=0.163773in,left,base]{\color{textcolor}\sffamily\fontsize{8.000000}{9.600000}\selectfont ECMP}%
\end{pgfscope}%
\begin{pgfscope}%
\pgfsetroundcap%
\pgfsetroundjoin%
\pgfsetlinewidth{1.505625pt}%
\definecolor{currentstroke}{rgb}{0.968627,0.505882,0.749020}%
\pgfsetstrokecolor{currentstroke}%
\pgfsetdash{}{0pt}%
\pgfpathmoveto{\pgfqpoint{2.798904in}{0.202662in}}%
\pgfpathlineto{\pgfqpoint{2.910015in}{0.202662in}}%
\pgfpathlineto{\pgfqpoint{3.021126in}{0.202662in}}%
\pgfusepath{stroke}%
\end{pgfscope}%
\begin{pgfscope}%
\definecolor{textcolor}{rgb}{0.150000,0.150000,0.150000}%
\pgfsetstrokecolor{textcolor}%
\pgfsetfillcolor{textcolor}%
\pgftext[x=3.110015in,y=0.163773in,left,base]{\color{textcolor}\sffamily\fontsize{8.000000}{9.600000}\selectfont SR}%
\end{pgfscope}%
\begin{pgfscope}%
\pgfsetroundcap%
\pgfsetroundjoin%
\pgfsetlinewidth{1.505625pt}%
\definecolor{currentstroke}{rgb}{0.650980,0.337255,0.156863}%
\pgfsetstrokecolor{currentstroke}%
\pgfsetdash{}{0pt}%
\pgfpathmoveto{\pgfqpoint{3.486588in}{0.202662in}}%
\pgfpathlineto{\pgfqpoint{3.597699in}{0.202662in}}%
\pgfpathlineto{\pgfqpoint{3.708811in}{0.202662in}}%
\pgfusepath{stroke}%
\end{pgfscope}%
\begin{pgfscope}%
\definecolor{textcolor}{rgb}{0.150000,0.150000,0.150000}%
\pgfsetstrokecolor{textcolor}%
\pgfsetfillcolor{textcolor}%
\pgftext[x=3.797699in,y=0.163773in,left,base]{\color{textcolor}\sffamily\fontsize{8.000000}{9.600000}\selectfont Best}%
\end{pgfscope}%
\begin{pgfscope}%
\pgfsetroundcap%
\pgfsetroundjoin%
\pgfsetlinewidth{1.505625pt}%
\definecolor{currentstroke}{rgb}{0.596078,0.305882,0.639216}%
\pgfsetstrokecolor{currentstroke}%
\pgfsetdash{}{0pt}%
\pgfpathmoveto{\pgfqpoint{4.242252in}{0.202662in}}%
\pgfpathlineto{\pgfqpoint{4.353363in}{0.202662in}}%
\pgfpathlineto{\pgfqpoint{4.464475in}{0.202662in}}%
\pgfusepath{stroke}%
\end{pgfscope}%
\begin{pgfscope}%
\definecolor{textcolor}{rgb}{0.150000,0.150000,0.150000}%
\pgfsetstrokecolor{textcolor}%
\pgfsetfillcolor{textcolor}%
\pgftext[x=4.553363in,y=0.163773in,left,base]{\color{textcolor}\sffamily\fontsize{8.000000}{9.600000}\selectfont Balanced ECMP}%
\end{pgfscope}%
\end{pgfpicture}%
\makeatother%
\endgroup%
    
    \begin{subfigure}[b]{0.49\linewidth}
     \centering
     \input{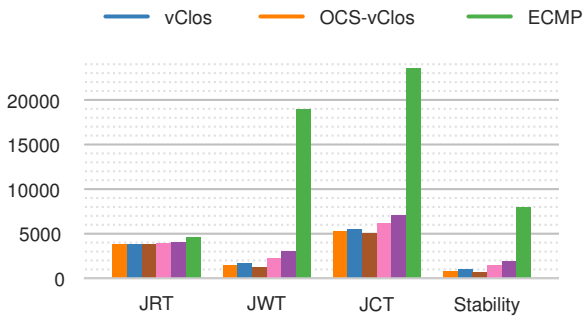}
     \caption{\textsc{Cluster512}}\label{fig:ind_512}
    \end{subfigure}
    \begin{subfigure}[b]{0.49\linewidth}
     \centering
     \input{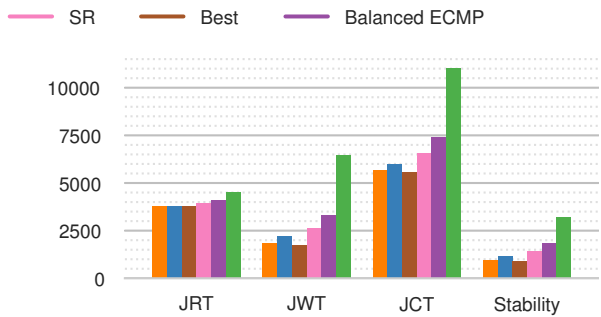}
     \caption{\textsc{Cluster2048}}\label{fig:ind_2048}
    \end{subfigure}
    
\caption{Key Performance Indicators Average on \textsc{Cluster512} and \textsc{Cluster2048}.}
\end{figure*}

\subsection{Baselines for Comparison}
We compare our algorithm \emph{vClos} and \emph{OCS-vClos} with \emph{Source Routing}(SR) and ECMP. We also tested two strategies in simulation experiment: i) Balanced ECMP(\emph{Balanced}): Beyond traditional ECMP \cite{Hopps2000AnalysisOA, Chiesa2014TrafficEW}, we assume that we can sense the current link load of the network, we randomly select the least congested ports to transmit instead of directly random ports. Its performance is better than traditional ECMP; ii) \emph{Best}: we assume that there is an electric switch with enough ports to connect all servers. In this case, network contention can be completely avoided. This is also the upper bound of performance among all strategies. We measure cluster performance using $Avg. JRT$, $Avg. JCT$, and $Avg. JWT$, and measure the $Stability$ of tasks by the average standard deviation of $JCT$ of DML with same parameter.

\subsection{Performance Analysis of large-scale clusters.}
Simulation experiment is conducted to analyse performance of large-scale clusters, where we set $\lambda=120s$ for \textsc{Cluster512} to keep the system in a steady state with task queuing. The results in Fig. \ref{fig:ind_512} shows that both vClos and OCS-vClos policy can reduce $Avg. JRT$ by 3.1\% compared with \emph{Source Routing} and 6.17\% compared with \emph{Balanced ECMP}, indicating that the elimination of network contention can reduce the training time of DML tasks. 

In Fig. \ref{fig:JWT}, \ref{fig:queue}, and \ref{fig:ind_512}, although $Avg. JRT$ only reduces by 3.18\% when using \emph{OCS-vClos} compared with \emph{SR}, $Avg. JWT$ is reduced by 65.65\%. The reason is that according to queuing theory, even a slight change in arrival rate may cause the system to enter a non-stationary state for a period of time, resulting in a significant increase in average waiting time. The introduction of OCS layer can reduce $Avg. JWT$ by 20.35\%, which shows reconfiguring network connections by OCS can effectively reduce network resource fragmentation. $Avg. JCT$ is only increased by 3.38\% compared with \emph{Best} when using \emph{OCS-vClos}, 
which indicates that the performance when using \emph{OCS-vClos} is very close to the theoretical upper limit.

Among all strategies, when using \emph{ECMP}, distribution of $JWT$ has a significant long tail as shown in Fig. \ref{fig:JRT}, indicating that hash-collision can cause significant performance fluctuations in certain tasks and affect user experience.  Although \emph{Source Routing} and \emph{Balanced ECMP} can effectively reduce hash collisions, it still cannot completely eliminate network contention. 

Fig. \ref{fig:stab_512} shows the $Stability$ of different kind of jobs under different strategies. The $Stability$ is mainly influenced by the $JRT$, and the small task and the task using 160GPU have lower $Stability$ due to the large variance of the running time. Some tasks, such as those using 96GPU, have high $Stability$ due to the small sample size. However, regardless of the type of task, the $Stability$ of the task is relatively low when using the ECMP strategy, which is consistent with the results of our testbed experiment.

\begin{table*}[htbp]
	\centering 
	\caption{Avg. JCT(s) of different strategy under different workload on Cluster512}
	\label{workload_oxc}
	\begin{tabular}{c|ccccccc}  
	    \toprule  
	    $\lambda(s)$&OCS-vClos&vClos&Best&SR&Balanced&ECMP&OCS-Releax\\
	    \midrule
		100& 7800.5 & 9799.6 & 7514.3 & 10083.5 & 12267.1 & 35032.9&24313.3\\  
		110& 6368.0 & 6981.5 & 6131.8 & 7449.6 & 8622.5 & 25055.6&13768.4\\   
		120& 5229.9 & 5550.6 & 5053.0 & 6228.6 & 7105.4 & 23545.4&11291.4\\   
		130& 4491.5 & 4710.4 & 4340.7 & 4955.6 & 5436.9 & 13118.7&7586.6\\   
		140& 4025.4 & 4105.1 & 3986.5 & 4209.3 & 4764.5 & 11395.6&5917.9\\   
		\bottomrule 
	\end{tabular}
\end{table*}

\subsection{Sensitivity to Cluster Size}
We also conducted simulations on a larger cluster, \textsc{Cluster2048}, which contains 2048 GPUs, and switches and servers of the same model as \textsc{Cluster512}. In order to keep the system in a reasonable steady state, $lambda$ is set to $15s$. Similarly to the case in  \textsc{Cluster512}, Fig.~\ref{fig:ind_2048} shows that \emph{OCS-vClos} can reduce $Avg. JRT$ by 1.59\% and $Avg. JWT$ by 64.91\% compared with \emph{Source Routing}. However, $Avg. JWT$ when using \emph{vClos} is only increased by 12.06\% compared with \emph{OCS-vClos}, which may be due to more available resources in larger-scale scenarios, resulting in a corresponding reduction in network resource fragmentation. This suggests that using \emph{vClos} in large-scale scenarios is sufficient to ensure task performance. 

\subsection{Sensitivity to workload}
In the simulation experiment, the task arrival time is randomly generated according to the Poisson distribution, and the $Stability$ of the queuing system is mainly determined by the arrival rate. Therefore, we try to change the task arrival rate by changing the parameters $\lambda$ of the Poisson distribution, and analyze the performance of vClos and OCS-vClos under different arrival rates. To be more specific, we choose $\lambda=100s,110s,120s,130s,140s$. Results in table \ref{workload_oxc} shows that the larger the $\lambda$, the more sensitive $Avg. JCT$ is to network contention, so it is more necessary to eliminate network contention in a clusters with high utilization. At the same time, when $\lambda$ is large, the advantage of \emph{OCS-vClos} relative to \emph{vClos} is more obvious, which may be because the probability of network resource fragmentation is also higher when the resource utilization rate is higher. What is more, results of \emph{OCS-Releax} indicate that relaxing the locality limit may greatly increase training time. Under various workload conditions, the performance gap between \emph{OCS-vClos} and \emph{Best} remains within 4\%, which verifies that OCS-vClos is not sensitive to workload. 

\subsection{The impact of job scheduler}
\label{job_scheduler}
Many previous articles mentioned the impact of job scheduler, our proposed algorithm mainly focuses on the invocation of network resources, and we wondered whether our framework still powerful under different job schedulers. We implemented fewest GPU first(FF), early deadline first(EDF) in simulation experiments. The results in Table \ref{scheduler} show that when the scheduling algorithm performs better, the sensitivity of $Avg. JCT$ to network contention and network fragmentation becomes lower, this is because well performing scheduling algorithms can reduce $Avg. JWT$. But the \emph{vClos} strategy still outperforms other strategies.

\begin{table}[htbp]
	\centering 
	\caption{Avg. JCT(s) under different job scheduler}
	\label{scheduler}
	\setlength{\tabcolsep}{0.5mm}
	\begin{tabular}{c|cccccc}  
	    \toprule
	    Sched.&OCS-vClos&vClos&Best&SR&Balanced&ECMP\\
	    \midrule  
		FIFO& 5229.9 & 5550.6 & 5053.0 & 6228.6 & 7105.4 & 23545.4\\ 
		EDF& 4198.5 & 4401.4 & 4176.5 & 4685.5 & 5020.9 & 7658.6\\  
		FF& 4081.6 & 4137.2 & 4029.9 & 4250.5 & 4607.8 & 6667.6\\   
		\bottomrule
	\end{tabular}
\end{table}

\subsection{The impact of Job Distribution}
In recent years, with the increasing demand for computing power in tasks, the number of GPUs required for a single task has also been increasing. To verify the scalability of \emph{vClos}, we conducted experiments by regenerating a dataset based on the data in the TPUv4 paper. The experimental results are shown in Table~\ref{job_distribution}, and it can be seen that the performance of \emph{vClos} is same as \emph{OCS-vClos}. This may be because the tasks in TPUv4 are mainly large tasks, so the occupation of network resources is relatively regular, resulting in less fragmentation of network resources, which also implies that as the number of tasks increases, \emph{vClos} is sufficient to ensure task performance.

\begin{table}[htbp]
	\centering 
	\caption{Avg. JRT(s),Avg. JWT(s) and Avg. JCT(s) under different job distribution}
	\label{job_distribution}
	\setlength{\tabcolsep}{0.35mm}
	\begin{tabular}{c|cccccc}  
	    \toprule
	   Crit.&OCS-vClos&vClos&Best&SR&Balanced&ECMP\\
	    \midrule  
		JRT& 3856.1 & 3856.1 & 3856.1 & 3858.0 & 3861.6 & 4968.8\\ 
		JWT& 8101.9& 8102.8 & 8101.9 & 8115.7 & 8160.2 & 17839.8\\  
		JCT& 11958.1 & 11958.9 & 11958.1 & 11973.7 & 12021.8 & 22808.7\\   
		\bottomrule
	\end{tabular}
\end{table}

\section{Relative Work}
How to reduce the impact of network overhead on performance has been a hot topic recent years. There exist many schemes. The first scheme is to use parameter compression to reduce the amount of communication data during parameter synchronization.  However, lossy compression, such as gradient thinning and gradient quantization \cite{20183LC}\cite{2021Gradient}, may reduce the accuracy of the final AI model. 

The second scheme is to adopt the pipeline mode \cite{jiang2020unified,peng2019generic,wang2020geryon,shi2021exploiting,romero2022accelerating,huang2019gpipe,narayanan2019pipedream}, that is, divide the Al model into multiple groups of tensors, so that the tensor group that has completed the calculation can start communication ahead of time while calculating a group of tensors, and the communication time can be hidden in the calculation process. However, those work does not care about conflict network environment on a multi-tenant cluster. 

The third solution is to reduce the impact of communication bottleneck between nodes/servers on distributed training tasks by reasonably deploying AI distributed training tasks \cite{xiao2018gandiva,narayanan2020heterogeneity,zhao2020hived,qiao2021pollux,jeon2019analysis,hu2021characterization,weng2022mlaas}. But these work generally treat the network as a black box and fail to take into account the potential problems of the network itself.

The fourth solution is to reduce communication overhead by improving the bandwidth utilization. \cite{mai2015optimizing,yang2021traffic,qi2021rationing,viswanathan2020network} reduce the impact of network contention by traffic management, \cite{wang2022topoopt} proposes intelligent cluster network-efficient scheduling methods by resource estimation, but they do not totally eliminate network contention. \cite{2018Highly,cho2019blueconnect} are designed to avoid the low bandwidth between servers slowing down the entire communication process by designing a multi-stage communication mode. BLink \cite{wang2020blink} and Plink \cite{luo2020plink} aim at generating the optimal communication scheme for distributed training through bandwidth detection. EFLOPS \cite{dong2020eflops} try to avoid network contention of a single task by jointly optimizing network topology and AllReduce communication in distributed training. These works help to improve the network bandwidth utilization efficiency of Al distributed training tasks, but cannot fundamentally eliminate the bandwidth bottleneck in a multi-tenant GPU cluster.

\section{Discussion}
\subsection{Stability of MEMS-OXC}
The MEMS-OXC we used in the experiment is a first-generation prototype of a certain company, and it is very sensitive to noise. When conduct testbed experiment, we often encounter network instability, which leads to frequent disconnection of RDMA connections. To alleviate this situation, we add soundproofing cotton to MEMS-OXC and regularly adjust the lenses to ensure signal stability. In future work, we will try to use a more reliable MEMS OXC.

\subsection{Low communication efficiency of hierarchical algorithms}
Although we realize HD and Hierarchical Ring in NCCL, the communication efficiency of these two hierarchical algorithms is relatively slow. We found that the reason for this phenomenon is that NCCL sends data based on a block during each communication, with a block size of 4MB by default. However, in the layered algorithm, our point-to-point communication volume is smaller each time, which can lead to a smaller actual block size and affect communication efficiency. We will further explore this phenomenon in future work.


\section{Conclusion}
In this paper, we first propose a routing strategy suitable for AI training, and then propose \emph{vClos} to isolate traffic between tasks to eliminate network contention in a multi-tenant cluster. Then we propose \emph{OCS-vClos}, which further reduces fragmentation of resources through network configuration. Our testbed and simulation experiments demonstrate that the proposed strategy can effectively reduce $Avg. JRT$, $Avg. JCT$, and $Avg. JWT$ and improve task $Stability$ to enhance user experience.

\clearpage
\bibliography{bibliography.bib}

\begin{thebibliography}{10}

\bibitem{abadi2016tensorflow}
Mart{\'\i}n Abadi.
\newblock Tensorflow: learning functions at scale.
\newblock In {\em Proceedings of the 21st ACM SIGPLAN International Conference
  on Functional Programming}, pages 1--1, 2016.

\bibitem{awan2019scalable}
Ammar~Ahmad Awan, Jereon B{\'e}dorf, Ching-Hsiang Chu, Hari Subramoni, and
  Dhabaleswar~K Panda.
\newblock Scalable distributed dnn training using tensorflow and cuda-aware
  mpi: Characterization, designs, and performance evaluation.
\newblock In {\em 2019 19th IEEE/ACM International Symposium on Cluster, Cloud
  and Grid Computing (CCGRID)}, pages 498--507. IEEE, 2019.

\bibitem{2021Gradient}
Youhui Bai, Cheng Li, Quan Zhou, Jun Yi, Ping Gong, Feng Yan, Ruichuan Chen,
  and Yinlong Xu.
\newblock Gradient compression supercharged high-performance data parallel dnn
  training.
\newblock In {\em Proceedings of the ACM SIGOPS 28th Symposium on Operating
  Systems Principles}, SOSP '21, page 359–375, New York, NY, USA, 2021.
  Association for Computing Machinery.

\bibitem{budzianowski2019hello}
Pawel Budzianowski and Ivan Vulic.
\newblock Hello, it's {GPT-2} - how can {I} help you? towards the use of
  pretrained language models for task-oriented dialogue systems.
\newblock {\em CoRR}, abs/1907.05774, 2019.

\bibitem{RapidNetSim}
Peirui Cao.
\newblock Rapidnetsim.
\newblock \url{https://github.com/caopeirui/rapidnetsim/}.

\bibitem{chen2015mxnet}
Tianqi Chen, Mu~Li, Yutian Li, Min Lin, Naiyan Wang, Minjie Wang, Tianjun Xiao,
  Bing Xu, Chiyuan Zhang, and Zheng Zhang.
\newblock Mxnet: {A} flexible and efficient machine learning library for
  heterogeneous distributed systems.
\newblock {\em CoRR}, abs/1512.01274, 2015.

\bibitem{cheng2019bandwidth}
Zehua Cheng and Zhenghua Xu.
\newblock Bandwidth reduction using importance weighted pruning on ring
  allreduce.
\newblock {\em CoRR}, abs/1901.01544, 2019.

\bibitem{Chiesa2014TrafficEW}
Marco Chiesa, Guy Kindler, and Michael Schapira.
\newblock Traffic engineering with equal-cost-multipath: An algorithmic
  perspective.
\newblock In {\em IEEE INFOCOM 2014 - IEEE Conference on Computer
  Communications}, pages 1590--1598, 2014.

\bibitem{cho2019blueconnect}
Minsik Cho, Ulrich Finkler, David Kung, and Hillery Hunter.
\newblock Blueconnect: Decomposing all-reduce for deep learning on
  heterogeneous network hierarchy.
\newblock volume~1, pages 241--251, 2019.

\bibitem{chu2020nv}
Ching-Hsiang Chu, Pouya Kousha, Ammar~Ahmad Awan, Kawthar~Shafie Khorassani,
  Hari Subramoni, and Dhabaleswar K. (D~K) Panda.
\newblock Nv-group: Link-efficient reduction for distributed deep learning on
  modern dense gpu systems.
\newblock In {\em Proceedings of the 34th ACM International Conference on
  Supercomputing}, ICS '20, New York, NY, USA, 2020. Association for Computing
  Machinery.

\bibitem{devlin2018bert}
Jacob Devlin, Ming{-}Wei Chang, Kenton Lee, and Kristina Toutanova.
\newblock {BERT:} pre-training of deep bidirectional transformers for language
  understanding.
\newblock {\em CoRR}, abs/1810.04805, 2018.

\bibitem{dong2020eflops}
Jianbo Dong, Zheng Cao, Tao Zhang, Jianxi Ye, Shaochuang Wang, Fei Feng,
  Li~Zhao, Xiaoyong Liu, Liuyihan Song, Liwei Peng, Yiqun Guo, Xiaowei Jiang,
  Lingbo Tang, Yin Du, Yingya Zhang, Pan Pan, and Yuan Xie.
\newblock Eflops: Algorithm and system co-design for a high performance
  distributed training platform.
\newblock In {\em 2020 IEEE International Symposium on High Performance
  Computer Architecture (HPCA)}, pages 610--622, 2020.

\bibitem{dryden2018aluminum}
Nikoli Dryden, Naoya Maruyama, Tim Moon, Tom Benson, Andy Yoo, Marc Snir, and
  Brian Van~Essen.
\newblock Aluminum: An asynchronous, gpu-aware communication library optimized
  for large-scale training of deep neural networks on hpc systems.
\newblock In {\em 2018 IEEE/ACM Machine Learning in HPC Environments (MLHPC)},
  pages 1--13, 2018.

\bibitem{fei2021efficient}
Jiawei Fei, Chen-Yu Ho, Atal~N. Sahu, Marco Canini, and Amedeo Sapio.
\newblock Efficient sparse collective communication and its application to
  accelerate distributed deep learning.
\newblock In {\em Proceedings of the 2021 ACM SIGCOMM 2021 Conference}, SIGCOMM
  '21, page 676–691, New York, NY, USA, 2021. Association for Computing
  Machinery.

\bibitem{floridi2020gpt}
Luciano Floridi and Massimo Chiriatti.
\newblock Gpt-3: Its nature, scope, limits, and consequences.
\newblock {\em Minds and Machines}, 30(4):681--694, 2020.

\bibitem{web:demand_doble}
Karen Hao.
\newblock The computing power needed to train ai is now rising seven times
  faster than ever before.
\newblock Accessed 6 May 2019, 2019.

\bibitem{Hopps2000AnalysisOA}
Christian~E. Hopps.
\newblock Analysis of an equal-cost multi-path algorithm.
\newblock RFC 2992, RFC Editor, 2000.

\bibitem{hu2021characterization}
Qinghao Hu, Peng Sun, Shengen Yan, Yonggang Wen, and Tianwei Zhang.
\newblock Characterization and prediction of deep learning workloads in
  large-scale gpu datacenters.
\newblock In {\em SC21: International Conference for High Performance
  Computing, Networking, Storage and Analysis}, pages 1--15, 2021.

\bibitem{huang2019gpipe}
Yanping Huang, Youlong Cheng, Ankur Bapna, Orhan Firat, Mia~Xu Chen, Dehao
  Chen, HyoukJoong Lee, Jiquan Ngiam, Quoc~V. Le, Yonghui Wu, and Zhifeng Chen.
\newblock Gpipe: Efficient training of giant neural networks using pipeline
  parallelism.
\newblock In {\em Proceedings of the 33rd International Conference on Neural
  Information Processing Systems}, Red Hook, NY, USA, 2019. Curran Associates
  Inc.

\bibitem{jeaugey2019massively}
Sylvain Jeaugey.
\newblock Massively scale your deep learning training with nccl 2.4.
\newblock {\em 2019, nVIDIA Developer Blog}, 2019.

\bibitem{jeon2019analysis}
Myeongjae Jeon, Shivaram Venkataraman, Amar Phanishayee, unjie Qian, Wencong
  Xiao, and Fan Yang.
\newblock Analysis of large-scale multi-tenant gpu clusters for dnn training
  workloads.
\newblock In {\em Proceedings of the 2019 USENIX Conference on Usenix Annual
  Technical Conference}, USENIX ATC '19, page 947–960, USA, 2019. USENIX
  Association.

\bibitem{2018Highly}
Xianyan Jia, Shutao Song, Wei He, Yangzihao Wang, Haidong Rong, Feihu Zhou,
  Liqiang Xie, Zhenyu Guo, Yuanzhou Yang, Liwei Yu, Tiegang Chen, Guangxiao Hu,
  Shaohuai Shi, and Xiaowen Chu.
\newblock Highly scalable deep learning training system with mixed-precision:
  Training imagenet in four minutes.
\newblock {\em CoRR}, abs/1807.11205, 2018.

\bibitem{jia2018highly}
Xianyan Jia, Shutao Song, Wei He, Yangzihao Wang, Haidong Rong, Feihu Zhou,
  Liqiang Xie, Zhenyu Guo, Yuanzhou Yang, Liwei Yu, et~al.
\newblock Highly scalable deep learning training system with mixed-precision:
  Training imagenet in four minutes.
\newblock {\em arXiv preprint arXiv:1807.11205}, 2018.

\bibitem{jiang2020unified}
Yimin Jiang, Yibo Zhu, Chang Lan, Bairen Yi, Yong Cui, and Chuanxiong Guo.
\newblock A unified architecture for accelerating distributed {DNN} training in
  heterogeneous {GPU/CPU} clusters.
\newblock In {\em 14th USENIX Symposium on Operating Systems Design and
  Implementation (OSDI 20)}, pages 463--479. USENIX Association, November 2020.

\bibitem{lecun1988theoretical}
Yann LeCun, D~Touresky, G~Hinton, and T~Sejnowski.
\newblock A theoretical framework for back-propagation.
\newblock In {\em Proceedings of the 1988 connectionist models summer school},
  volume~1, pages 21--28, 1988.

\bibitem{20183LC}
Hyeontaek Lim, David~G. Andersen, and Michael Kaminsky.
\newblock 3lc: Lightweight and effective traffic compression for distributed
  machine learning.
\newblock {\em CoRR}, abs/1802.07389, 2018.

\bibitem{luo2020plink}
Liang Luo, Peter West, Arvind Krishnamurthy, Luis Ceze, and Jacob Nelson.
\newblock Plink: Discovering and exploiting locality for accelerated
  distributed training on the public cloud.
\newblock In {\em MLSys}, 2020.

\bibitem{mai2015optimizing}
Luo Mai, Chuntao Hong, and Paolo Costa.
\newblock Optimizing network performance in distributed machine learning.
\newblock In {\em 7th USENIX Workshop on Hot Topics in Cloud Computing
  (HotCloud 15)}, 2015.

\bibitem{narayanan2019pipedream}
Deepak Narayanan, Aaron Harlap, Amar Phanishayee, Vivek Seshadri, Nikhil~R
  Devanur, Gregory~R Ganger, Phillip~B Gibbons, and Matei Zaharia.
\newblock Pipedream: generalized pipeline parallelism for dnn training.
\newblock In {\em Proceedings of the 27th ACM Symposium on Operating Systems
  Principles}, pages 1--15, 2019.

\bibitem{narayanan2020heterogeneity}
Deepak Narayanan, Keshav Santhanam, Fiodar Kazhamiaka, Amar Phanishayee, and
  Matei Zaharia.
\newblock Heterogeneity-aware cluster scheduling policies for deep learning
  workloads.
\newblock In {\em OSDI}, 2020.

\bibitem{DGX2023}
{NVIDIA}.
\newblock {NVIDIA DGX SuperPOD}: Next generation scalable infrastructure for ai
  leadership.
\newblock [Online]. Available: \url{https://shorturl.at/mxzB2}.
\newblock Accessed Dec., 2023.

\bibitem{paszke2019pytorch}
Adam Paszke, Sam Gross, Francisco Massa, Adam Lerer, James Bradbury, Gregory
  Chanan, Trevor Killeen, Zeming Lin, Natalia Gimelshein, Luca Antiga, et~al.
\newblock Pytorch: An imperative style, high-performance deep learning library.
\newblock {\em Advances in neural information processing systems}, 32, 2019.

\bibitem{peng2019generic}
Yanghua Peng, Yibo Zhu, Yangrui Chen, Yixin Bao, Bairen Yi, Chang Lan, Chuan
  Wu, and Chuanxiong Guo.
\newblock A generic communication scheduler for distributed dnn training
  acceleration.
\newblock In {\em Proceedings of the 27th ACM Symposium on Operating Systems
  Principles}, pages 16--29, 2019.

\bibitem{qi2021rationing}
Qiang Qi, Fei Xu, Li~Chen, and Zhi Zhou.
\newblock Rationing bandwidth resources for mitigating network resource
  contention in distributed dnn training clusters.
\newblock {\em CCF Transactions on High Performance Computing}, 3(2):171--185,
  2021.

\bibitem{qiao2021pollux}
Aurick Qiao, Sang~Keun Choe, Suhas~Jayaram Subramanya, Willie Neiswanger,
  Qirong Ho, Hao Zhang, Gregory~R. Ganger, and Eric~P. Xing.
\newblock Pollux: Co-adaptive cluster scheduling for goodput-optimized deep
  learning.
\newblock In {\em OSDI}, 2021.

\bibitem{2021Breakfast}
D.~Raghavan, P.~A. Levis, M.~A. Zaharia, and I.~Zhang.
\newblock Breakfast of champions: towards zero-copy serialization with nic
  scatter-gather.
\newblock In {\em Proceedings of the Workshop on Hot Topics in Operating
  Systems}, 2021.

\bibitem{rashidi2022themis}
Saeed Rashidi, William Won, Sudarshan Srinivasan, Srinivas Sridharan, and
  Tushar Krishna.
\newblock Themis: A network bandwidth-aware collective scheduling policy for
  distributed training of dl models.
\newblock In {\em Proceedings of the 49th Annual International Symposium on
  Computer Architecture}, pages 581--596, 2022.

\bibitem{romero2022accelerating}
Joshua Romero, Junqi Yin, Nouamane Laanait, Bing Xie, M~Todd Young, Sean
  Treichler, Vitalii Starchenko, Albina Borisevich, Alex Sergeev, and Michael
  Matheson.
\newblock Accelerating collective communication in data parallel training
  across deep learning frameworks.
\newblock In {\em 19th USENIX Symposium on Networked Systems Design and
  Implementation (NSDI 22)}, pages 1027--1040, 2022.

\bibitem{2016Generalisation}
Martin Ruefenacht, Mark Bull, and Stephen Booth.
\newblock Generalisation of recursive doubling for allreduce.
\newblock In {\em European Mpi Users Group Meeting}, pages 23--31, 2016.

\bibitem{web:full_use_GPU}
Ram Sagar.
\newblock How to take full advantage of gpus in large language models.
\newblock Accessed 6 May 2021. \url{https://rb.gy/tr4hu}, 2021.

\bibitem{sanghoon2023logical}
Jo~Sanghoon, Hyojun Son, and John Kim.
\newblock Logical/physical topology-aware collective communication in deep
  learning training.
\newblock In {\em 2023 IEEE International Symposium on High-Performance
  Computer Architecture (HPCA)}, pages 56--68. IEEE, 2023.

\bibitem{sergeev2018horovod}
Alexander Sergeev and Mike Del~Balso.
\newblock Horovod: fast and easy distributed deep learning in tensorflow.
\newblock {\em arXiv preprint arXiv:1802.05799}, 2018.

\bibitem{shi2021exploiting}
Shaohuai Shi, Xiaowen Chu, and Bo~Li.
\newblock Exploiting simultaneous communications to accelerate data parallel
  distributed deep learning.
\newblock In {\em IEEE INFOCOM 2021-IEEE Conference on Computer
  Communications}, pages 1--10. IEEE, 2021.

\bibitem{thangakrishnan2020herring}
Indu Thangakrishnan, Derya Cavdar, Can Karakus, Piyush Ghai, Yauheni
  Selivonchyk, and Cory Pruce.
\newblock Herring: Rethinking the parameter server at scale for the cloud.
\newblock In {\em SC20: International Conference for High Performance
  Computing, Networking, Storage and Analysis}, pages 1--13. IEEE, 2020.

\bibitem{tong2021study}
Zhihao Tong, Ning Du, Xiaobo Song, and Xiaoli Wang.
\newblock Study on mindspore deep learning framework.
\newblock In {\em 2021 17th International Conference on Computational
  Intelligence and Security (CIS)}, pages 183--186. IEEE, 2021.

\bibitem{ueno2019exhaustive}
Yuichiro Ueno and Rio Yokota.
\newblock Exhaustive study of hierarchical allreduce patterns for large
  messages between gpus.
\newblock In {\em 2019 19th IEEE/ACM International Symposium on Cluster, Cloud
  and Grid Computing (CCGRID)}, pages 430--439. IEEE, 2019.

\bibitem{vaswani2017attention}
Ashish Vaswani, Noam Shazeer, Niki Parmar, Jakob Uszkoreit, Llion Jones,
  Aidan~N Gomez, {\L}ukasz Kaiser, and Illia Polosukhin.
\newblock Attention is all you need.
\newblock {\em Advances in neural information processing systems}, 30, 2017.

\bibitem{viswanathan2020network}
Raajay Viswanathan, Arjun Balasubramanian, and Aditya Akella.
\newblock Network-accelerated distributed machine learning for multi-tenant
  settings.
\newblock In {\em Proceedings of the 11th ACM Symposium on Cloud Computing},
  pages 447--461, 2020.

\bibitem{wang2020blink}
Guanhua Wang, Shivaram Venkataraman, Amar Phanishayee, Nikhil Devanur, Jorgen
  Thelin, and Ion Stoica.
\newblock Blink: Fast and generic collectives for distributed ml.
\newblock {\em Proceedings of Machine Learning and Systems}, 2:172--186, 2020.

\bibitem{wang2020geryon}
Shuai Wang, Dan Li, and Jinkun Geng.
\newblock Geryon: Accelerating distributed cnn training by network-level flow
  scheduling.
\newblock In {\em IEEE INFOCOM 2020-IEEE Conference on Computer
  Communications}, pages 1678--1687. IEEE, 2020.

\bibitem{wang2022topoopt}
Weiyang Wang, Moein Khazraee, Zhizhen Zhong, Zhijao Jia, Dheevatsa Mudigere,
  Ying Zhang, Anthony Kewitsch, and Manya Ghobadi.
\newblock Topoopt: Optimizing the network topology for distributed dnn
  training.
\newblock {\em arXiv preprint arXiv:2202.00433}, 2022.

\bibitem{weng2022mlaas}
Qizhen Weng, Wencong Xiao, Yinghao Yu, Wei Wang, Cheng Wang, Jian He, Yong Li,
  Liping Zhang, Wei Lin, and Yu~Ding.
\newblock $\{$MLaaS$\}$ in the wild: Workload analysis and scheduling in
  $\{$Large-Scale$\}$ heterogeneous $\{$GPU$\}$ clusters.
\newblock In {\em 19th USENIX Symposium on Networked Systems Design and
  Implementation (NSDI 22)}, pages 945--960, 2022.

\bibitem{wu2023sip}
Zhenguo Wu, Liang~Yuan Dai, Ziyi Zhu, Asher Novick, Madeleine Glick, and Keren
  Bergman.
\newblock Sip architecture for accelerating collective communication in
  distributed deep learning.
\newblock In {\em 2023 Optical Fiber Communications Conference and Exhibition
  (OFC)}, pages 1--3. IEEE, 2023.

\bibitem{xiao2018gandiva}
Wencong Xiao, Romil Bhardwaj, Ramachandran Ramjee, Muthian Sivathanu, Nipun
  Kwatra, Zhenhua Han, Pratyush Patel, Xuan Peng, Hanyu Zhao, Quanlu Zhang, Fan
  Yang, and Lidong Zhou.
\newblock Gandiva: Introspective cluster scheduling for deep learning.
\newblock In {\em OSDI}, 2018.

\bibitem{xu2022ring}
Zimu Xu, Wei Tian, Yingxin Liu, Wanjun Ning, and Jingjin Wu.
\newblock A ring topology-based optimization approach for federated learning in
  d2d wireless networks.
\newblock {\em arXiv preprint arXiv:2212.02830}, 2022.

\bibitem{yang2021traffic}
Weihong Yang, Yang Qin, Zukai Jiang, and Xiaowen Chu.
\newblock Traffic management for distributed machine learning in rdma-enabled
  data center networks.
\newblock In {\em ICC 2021-IEEE International Conference on Communications},
  pages 1--6. IEEE, 2021.

\bibitem{zhang2020network}
Zhen Zhang, Chaokun Chang, Haibin Lin, Yida Wang, Raman Arora, and Xin Jin.
\newblock Is network the bottleneck of distributed training?
\newblock In {\em Proceedings of the Workshop on Network Meets AI \& ML}, pages
  8--13, 2020.

\bibitem{zhao2020hived}
Hanyu Zhao, Zhenhua Han, Zhi Yang, Quanlu Zhang, Fan Yang, Lidong Zhou, Mao
  Yang, Francis~CM Lau, Yuqi Wang, Yifan Xiong, et~al.
\newblock Hived: sharing a $\{$GPU$\}$ cluster for deep learning with
  guarantees.
\newblock In {\em OSDI}, 2020.

\bibitem{zhao2019minimal}
Shizhen Zhao, Rui Wang, Junlan Zhou, Joon Ong, Jeffrey~C Mogul, and Amin
  Vahdat.
\newblock Minimal rewiring: Efficient live expansion for clos data center
  networks.
\newblock In {\em 16th USENIX Symposium on Networked Systems Design and
  Implementation (NSDI 19)}, pages 221--234, 2019.

\end{thebibliography}
\bibliographystyle{plain}

\clearpage
\appendix
\section{Appendix}

\subsection{There is no sample path optimal solution to generate virtual Clos in leaf-spine architecture}
\begin{figure}[!htbp]
\centering
    \begin{subfigure}[b]{\linewidth}
     \centering
     \includegraphics[width=\textwidth]{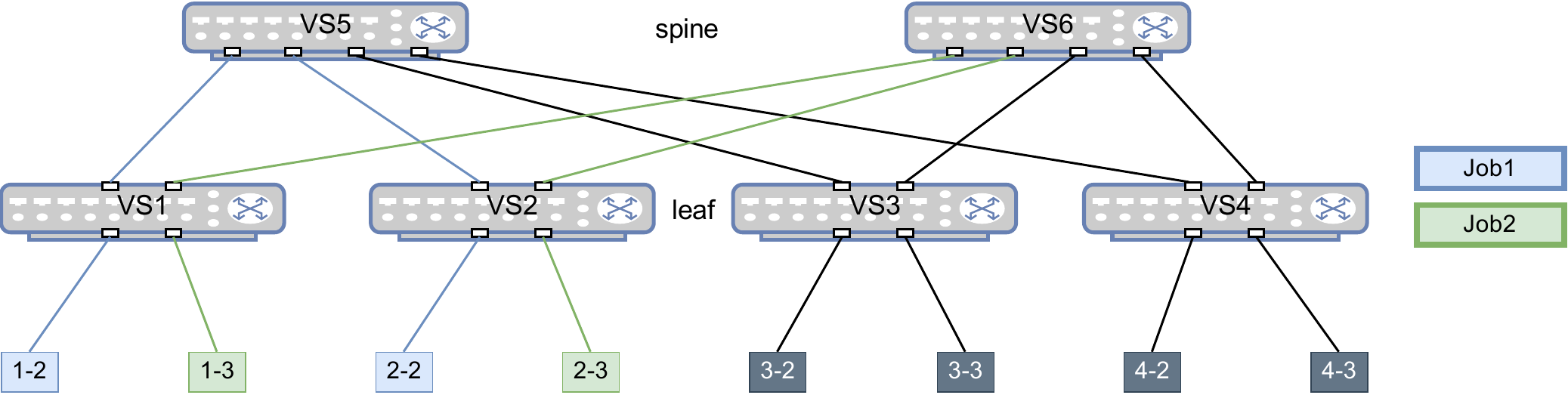}
     \caption{Initial Configuration}\label{fig:slot1}
 \end{subfigure}

    \begin{subfigure}[b]{\linewidth}
     \centering
     \includegraphics[width=\textwidth]{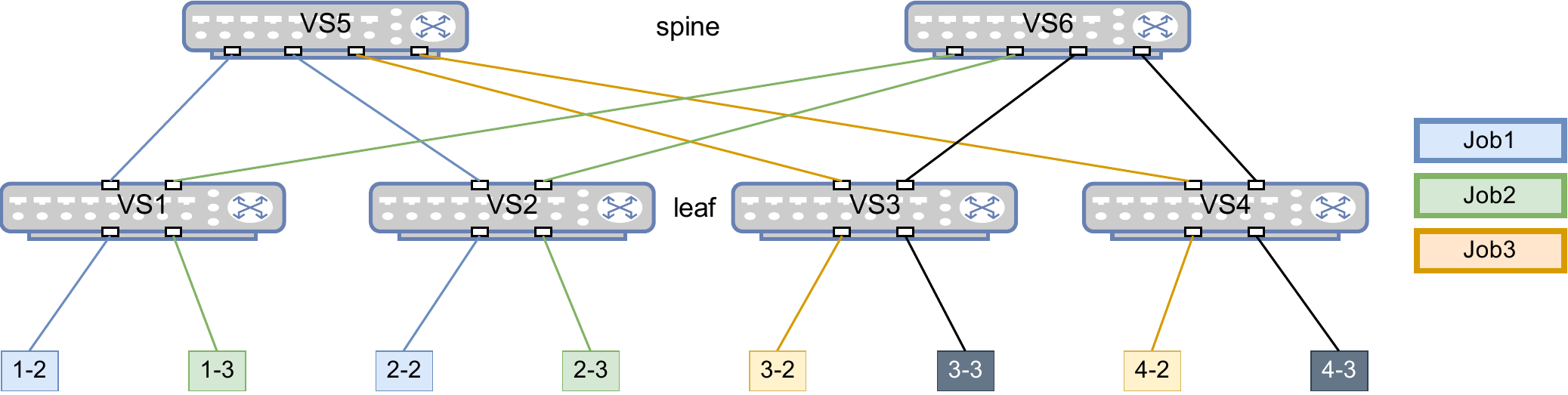}
     
     {\small\textsf{Job 1 finished first, and Job 4 cannot be allocated.}}
     \includegraphics[width=\textwidth]{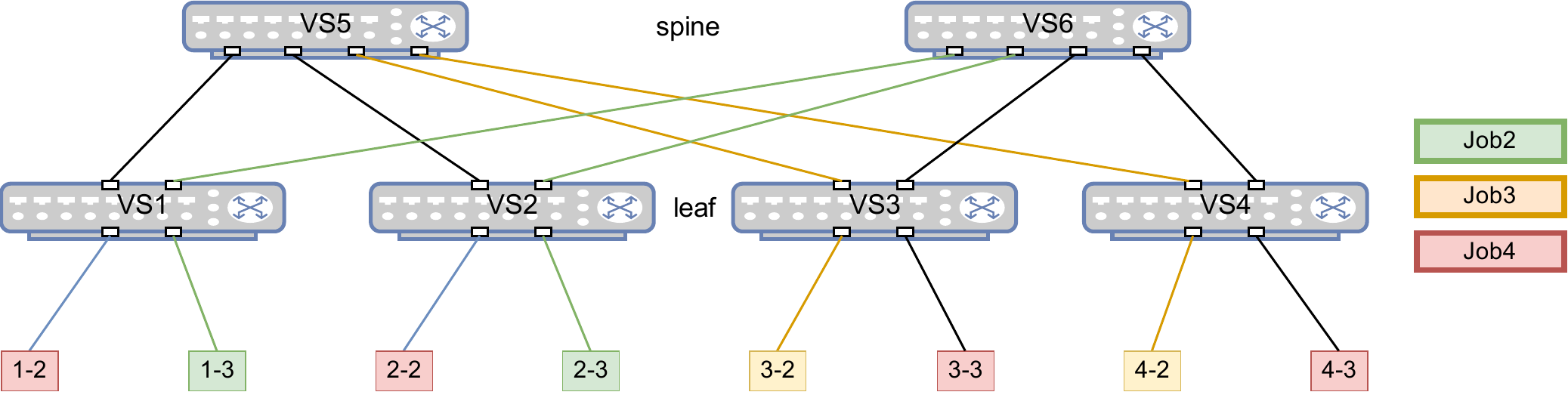}
     \caption{Allocation 1: through VS5}\label{fig:vs5}
 \end{subfigure}
 
 \begin{subfigure}[b]{\linewidth}
     \centering
     \includegraphics[width=\textwidth]{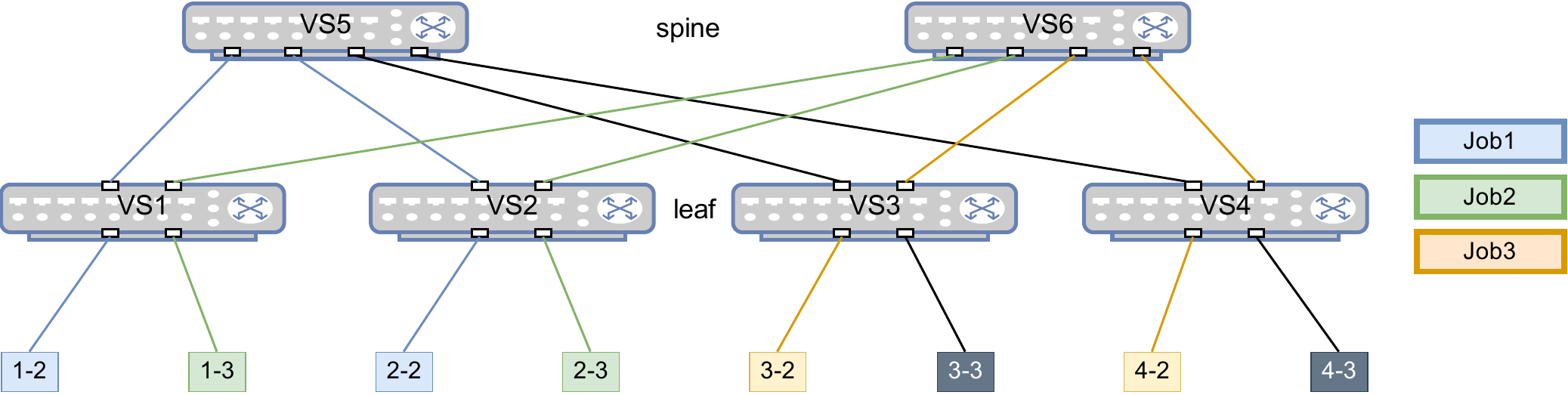}
     
     {\small\textsf{Job 2 finished first, and Job 4 cannot be allocated.}}
     \includegraphics[width=\textwidth]{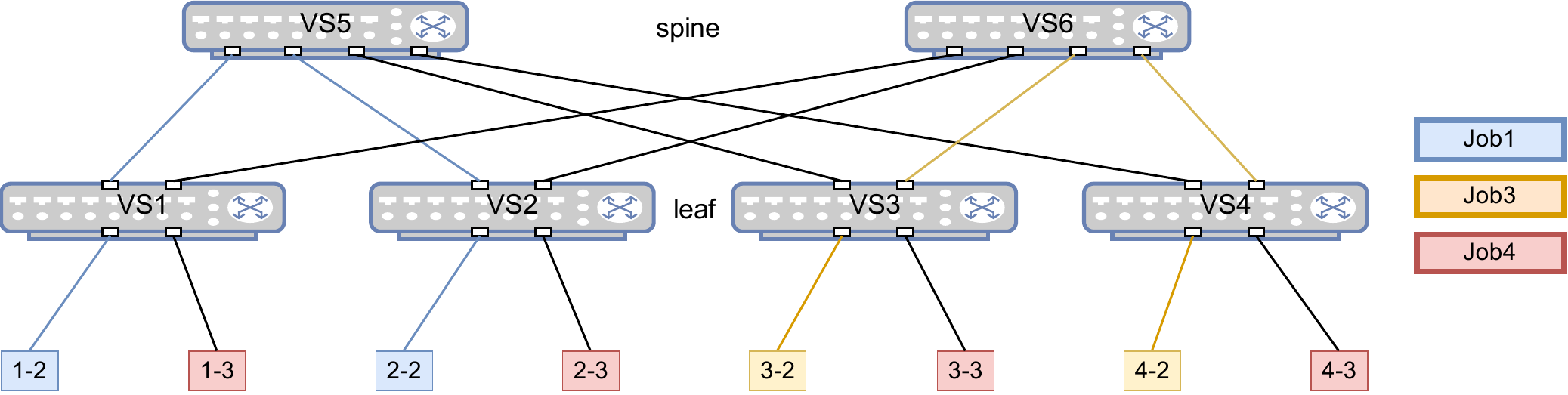}
     \caption{Allocation 2: through VS6}\label{fig:vs6}
 \end{subfigure}
 
\caption{A Simple Example}
\label{fig:conflict_fig}
\end{figure}

To illustrate that There is no sample path optimal solution to generate \emph{vClos} in leaf-spine architecture, we present a simple example depicted in Figure \ref{fig:conflict_fig}. Initially, the configuration is set as shown in Figure \ref{fig:slot1}, where Job 1 occupies GPU 1-2 and GPU 2-2, and Job 2 occupies GPU 1-3 and GPU 2-3. Subsequently, Job 3 arrives and occupies GPU 3-2 and GPU 4-2. At this point, two potential solutions emerge for generating a \emph{vClos} for Job 3. If we opt for the first solution as \ref{fig:vs5}, wherein Job 1 finishes in the next slot and releases the corresponding GPUs, a \emph{vClos} cannot be generated for Job 4. Likewise, the second solution, as shown in Figure \ref{fig:vs6}, encounters the same issue if Job 2 finishes in the following slot. Consequently, it becomes evident that no sample path optimal solution exists for this problem.

\subsection{ILP in vClos Stage 2}
In this section we introduce a \textsc{FindvClos} Algorithm to generate a \emph{vClos} with accessible time cost. $l_n^i$ and $s_m^i$ represent the number of virtual leaf and spine generated in $L_n$ and $S_m$ for $i$-th job, and $c_{n,m}$ represents whether the physical links between $L_n$ and $S_m$ will be used to form the virtual Clos. Our goal in vClos Generation Algorithm is to find the feasible solution of server resource parameter $r_n^i$ and network resource parameter $l_n^i$, $s_m^i$, $c_{n,m}$. For ease of reference, notations are summarized in Table \ref{table0} and Table \ref{tbl:notation_appendix_vclos}.


\begin{table}[!htb]
	\centering  
	\caption{Notations used in vClos Stage 2}\label{tbl:notation_appendix_vclos} 
	\begin{tabular}{c|l}  
		\toprule 
		$l_n^i$ & \begin{tabular}[c]{@{}l@{}}
		Num of virtual leaf in $L_n$ for $i$-th job\end{tabular}\\
		$s_m^i$ & \begin{tabular}[c]{@{}l@{}}
		Num of virtual spine in $S_m$ for $i$-th job\end{tabular}\\
	    $c_{n,m}$& \begin{tabular}[c]{@{}l@{}}
		Num of links to be used between between $L_n$ and $S_m$\end{tabular}\\
		\bottomrule
	\end{tabular}
\end{table}

Through experiments, it was found that most network fragmentation occurs because suitable leaf port resources cannot be found. Therefore, we first try to ensure that leaf port resources are occupied as neatly as possible, that is, the number of ports in each virtual leaf is as large as possible to a power of 2. The scheduler starts from the initial value of $l=max\{1,\frac{2^{\lfloor \log_{2}{N_i} \rfloor}}{S}\}$, $s$ = ($l=1$)? 0 : $\frac{N_i}{l}$, and tries every possible combination of p and q until it finds a suitable solution. For each given $l$ and $s$ vClos tries to solve a vClos-ILP which is deifined as Algorithm \ref{alg:find_vclos}.

\begin{algorithm}[!htbp]  
	\caption{\textsc{FindvClos} Algorithm}
	\LinesNumbered 
	\KwIn{Required job size $N_i$}
	\KwOut{$gpus$, $c_{n,m}$}
	$l \leftarrow \max\{1,\frac{2^{\lfloor \log_{2}{N_i} \rfloor}}{S}\}$\;
	$s \leftarrow$ ($l=1$)? 0 : $\frac{N_i}{l}$\;
	\For{$n \leftarrow 1$ \KwTo $S$}{
		$success, l_n^i, s_m^i, c_{n,m}$ = $vClosILP(l,s)$\;
		\If{$success$}{
		    \For{$n \leftarrow 1$ \KwTo $L$}{
		        $gpus.extend(L_n.ChooseGpu(T,\frac{l_n^i*s}{T}))$\;
		     }
			\Return $gpus,c_{n,m}$\;
		}
		$l \leftarrow l\times 2$\;
		$s \leftarrow \frac{N_i}{l}$\;
	}
	\Return $none, none$
\label{alg:find_vclos}
\end{algorithm}

\paragraph{vClos-ILP Constraints}
In \emph{vClos}, at most one virtual leaf/virtual spine can be generated from each leaf/spine for one training task, and the number of generated virtual leafs/spines should meet the number requirement. i.e.,
\begin{equation}\label{eq:1}
\begin{cases*}
\sum_n l_n^i = l \\
\sum_m s_m^i = s 
\end{cases*}
\end{equation}

By clos network definition, the number of uplinks must be equal to the number of virtual spine switch and the number of downlinks must be equal to the number of virtual leaf switch:
\begin{equation}\label{eq:6}
\begin{cases*}
\forall n, \sum_m c_{n,m}=s &if $l_n^i=1$\\
\forall m, \sum_n c_{n,m}=l &if $s_m^i=1$
\end{cases*}
\end{equation}


Recall that $C_{n,m}$ is the total number of unused link beween $L_n$ and $S_m$. Clearly, the link number to be used $C_{n,m}$ must be smaller than $l_n$ and $s_m$:
\begin{equation}\label{eq:3}
\forall n\forall m,  c_{n,m}\leq \min\{C_{n,m}, l_n^i, s_m^i\}
\end{equation}


Due to strict locality constraint, only idle servers can be used to generate \emph{vClos} and the number of to be used GPU in each virtual leaf should equal to the number of virtual spine:

\begin{equation}\label{eq:4}
\forall i, \begin{cases*}
r_n^i \leq R_n\\
r_n^i \geq l_n\\
T*r_n^i = s &if $l_n^i = 1$
\end{cases*}
\end{equation}


\paragraph{vClos-ILP Objective} By experience, the scheduler should look for solutions that have the least impact on idle leaf switches and spine switches, specifically, we use $RPN(S_m)$ defines the free ports number of $S_m$ before scheduling and $RSN(L_n)$ defines the free servers number of $L_n$ before scheduling. vClos chooses leafs with least free servers and spines with least free ports among all valid leafs/spines, i.e.,
\begin{equation}
\boxed{
\begin{aligned}
    \label{eq:5}
    &\min  \sum_m RPN(S_m)*s_m+ \sum_n {RSN(L_n)*T*l_n}\\
    \text{s.t. }& (\ref{eq:1}),(\ref{eq:6}),(\ref{eq:3}),(\ref{eq:4})
\end{aligned}
}
\end{equation}

\subsection{ILP in OCS-vClos Stage 3}
$l_n^i$ and $s_m^i$ represent the number of virtual leaf and spine generated in $L_n$ and $S_m$, and $c_{n,m}^k$ represents whether the physical links between $L_n$ and $S_m$ through $k$-th OCS $O_k$ will be used to form the \emph{OCS-vClos}. Recall in Equation \ref{eq:1}, each virtual leaf and virtual spine is connected by 1 link, but in OCS-vClos, we can reconfigure the link number between each leaf and spine as long as $\sum_{n,m,k} c_{n,m}^k=l_n^i*s_m^i$, unfortunately this constraint is nonlinear. To linearize this constraint, we define an intermediate variable $L_{n,a}^i$, where $\forall_{t<a} l_{n,t}^i = 1$ means $a$ virtual leafs are generated from $L_n$. Naturally, we have that $\sum_a L_{n,a}^i = l_n^i$. For ease of reference, notations are summarized in Table \ref{table3}.

\begin{table}[htbp]
	\centering  
	\setlength{\belowcaptionskip}{5pt}
	\caption{Notations used in OCS-vClos}\label{table3} 
	\begin{tabular}{c|l}  
		\toprule
		$l_n^i$& \begin{tabular}[c]{@{}l@{}}
		Num of virtual leaf in $L_n$ for $i$-th job\end{tabular}\\
		\hline 
		$s_m^i$& \begin{tabular}[c]{@{}l@{}}
		Num of virtual spine in $S_m$ for $i$-th job\end{tabular}\\ 
		\hline 
		$L_{n,a}^i$& \begin{tabular}[c]{@{}l@{}}
		A boolean intermediate variable to\\ represent virtual leaf generated in $L_n$ \end{tabular}\\ 
		\hline 
		$c_{n,a,m,k}$ &\begin{tabular}[c]{@{}l@{}}
		Whether the physical links between $L_{n,a}^i$ and\\ $s_m^i$ through OCS $O_k$ will be used \end{tabular}\\
		\bottomrule
        
	\end{tabular}
\end{table}

For given $l$ and $s$ OCS-vClos generates virtual Clos by solving following OCS-vClos-ILP:

\textbf{Physical Topology Constraints:} 
By clos network definition, the number of generated virtual leaf switches and spine switches should meet the required number,i.e., 

\begin{equation}\label{ocseq:1}
    \begin{cases}
    &\sum_a L_{n,a}^i = l_n^i \\
    &\sum_n\sum_a l_{n,a} = l \\
    &\sum_m s_{m} = s \\
    &\forall_{a>0}, L_{n,a}^i <= l_{n,a-1}^i \\
    \end{cases}
\end{equation}

  
Due to the locality constraint, the ports number of all virtual leaf switches occupies in $L_n$ should be equal to the ports number of all idle servers connected by $L_n$, i.e.,
\begin{equation}\label{ocseq:2}
    \forall n\forall a, L_{n,a}^i*s = T*r_n^i
\end{equation}

The number of links to be used should be equal to the number of GPUs the user requires, and the number of links between each $l_{n,a}$ and $S_{m}$ should be either 0 or the number of virtual spines in  $S_{m}$, i.e.

\begin{equation}\label{ocseq:3}
\begin{cases}
    \sum_{n,a,m,k} c_{n,a,m}^k = N^i \\
    \forall n\forall a \forall m, \sum_k c_{n,a,m}^k \leq \min\{L_{n,a}^i, s_{m}^i\}
\end{cases}
\end{equation}

All links out from $L_{n,a}$ should equal to the used port in $L_{n,a}$, i.e.

\begin{equation}\label{ocseq:4}
    \begin{cases}
        \forall n\forall a, \sum_{m,k} c_{n,a,m}^k = L_{n,a}^i*s \\
        \forall m  \sum_{n,a,k} c_{n,a,m}^k = s_{m}^i*l
    \end{cases}
\end{equation}

Link numbers between all $L_{n,a}^i$ and $S_m$ through $O_k$ can not larger than the potential links between $L_n$ and $S_m$ through $O_k$ , and the number of links between all generated $L_{n,a}^i$ and $O_k$ can be larger than the number of links between $L_n$ and $O_k$, similarly, the number of links between all generated $S_m$ and $O_k$ can be larger than the number of links between $S_m$ and $O_k$, i.e.

\begin{equation}\label{ocseq:5}
    \begin{cases}
        \forall a \forall k, \sum_{n,m} c_{n,a,m}^k \leq C_{n,m}^k\\
        \forall n \forall k, \sum_{m,a} c_{n,a,m}^k \leq C_n^k \\
        \forall m \forall k, \sum_{n,a} c_{n,a,m}^k \leq C_m^k
    \end{cases}
\end{equation}


\textbf{OCS-vClos-ILP Objective:} the objective function is same as it is in vClos, i.e.,

\begin{equation}\label{ocseq:6}
\boxed{
\begin{aligned}
    &\min  \sum_m RPN(S_m)*s_m+ \sum_n {RSN(L_n)*T*l_n}\\
    \text{s.t. } &(\ref{ocseq:2}),(\ref{ocseq:3}),(\ref{ocseq:4}),(\ref{ocseq:5})
\end{aligned}
}
\end{equation}

\begin{algorithm}[!htbp] 
\label{oxc_vclos_alg}
	\caption{\textsc{OCSFindClos} Algorithm}
	\LinesNumbered 
	\KwIn{Required job size $N^i$}
	\KwOut{$gpus$, $c_{n,m}^k$}
	$l \leftarrow \max\{1,\frac{2^{\lfloor \log_{2}{N_i} \rfloor}}{S}\}$\;
	$s \leftarrow$ ($l=1$)? 0 : $\frac{N_i}{l}$\;
	\While{$l \leq L$}{
		$success, l_n^i, s_m^i, c_{n,m}^k \leftarrow OCS-vClosILP(l,s)$
		\If{$success$}{
		    \For{$n \leftarrow 1$ \KwTo $S$}{
		        $ gpus.extend(L_n.ChooseGpu(T,\frac{l_n^i*s}{T})))$
		      }
			return $gpus, c_{n,m}^k$.
		}
		$l \leftarrow l\times 2$\;
		$s \leftarrow \frac{N_i}{l}$\;
	}
	
	return $none, none$.
\end{algorithm}






\end{document}